\documentclass[aps,eqsecnum,preprint,floats,epsf,epsfig,nofootinbib,letter]{revtex4}
\usepackage{epsfig}

\begin{document}
\def\be{\begin{eqnarray}}
\def\en{\end{eqnarray}}
\def\non{\nonumber}
\def\la{\langle}
\def\ra{\rangle}
\def\nc{N_c^{\rm eff}}
\def\vp{\varepsilon}
\def\drho{\bar\rho}
\def\deta{\bar\eta}
\def\CP{{\it CP}~}
\def\a{{\cal A}}
\def\B{{\cal B}}
\def\c{{\cal C}}
\def\d{{\cal D}}
\def\e{{\cal E}}
\def\p{{\cal P}}
\def\t{{\cal T}}
\def\up{\uparrow}
\def\dw{\downarrow}
\def\vma{{_{V-A}}}
\def\vpa{{_{V+A}}}
\def\smp{{_{S-P}}}
\def\spp{{_{S+P}}}
\def\J{{J/\psi}}
\def\ov{\overline}
\def\Lqcd{{\Lambda_{\rm QCD}}}
\def\pr{{Phys. Rev.}~}
\def\prl{{Phys. Rev. Lett.}~}
\def\pl{{Phys. Lett.}~}
\def\np{{Nucl. Phys.}~}
\def\zp{{Z. Phys.}~}
\def\lsim{ {\ \lower-1.2pt\vbox{\hbox{\rlap{$<$}\lower5pt\vbox{\hbox{$\sim$}
}}}\ } }
\def\gsim{ {\ \lower-1.2pt\vbox{\hbox{\rlap{$>$}\lower5pt\vbox{\hbox{$\sim$}
}}}\ } }

\font\el=cmbx10 scaled \magstep2{\obeylines\hfill November, 2005}

\vskip 1.5 cm

\centerline{\large\bf Charmless hadronic $B$ decays involving
scalar mesons:}
 \centerline{\large\bf Implications to the nature of light
scalar mesons}
\bigskip
\centerline{\bf Hai-Yang Cheng$^{1}$, Chun-Khiang Chua$^{1}$ and
Kwei-Chou Yang$^{2}$}
\medskip
\centerline{$^1$ Institute of Physics, Academia Sinica}
\centerline{Taipei, Taiwan 115, Republic of China}
\medskip
\centerline{$^2$ Department of Physics, Chung Yuan Christian
University} \centerline{Chung-Li, Taiwan 320, Republic of China}
\bigskip
\bigskip
\bigskip
\bigskip
\centerline{\bf Abstract}
\bigskip
\small
 The hadronic charmless $B$ decays into a scalar meson and a pseudoscalar
meson are studied within the framework of QCD factorization. Based
on the QCD sum rule method, we have derived the leading-twist
light-cone distribution amplitudes of scalar mesons and their
decay constants. Although the light scalar mesons $f_0(980)$ and
$a_0(980)$ are widely perceived as primarily the four-quark bound
states, in practice it is difficult to make quantitative
predictions based on the four-quark picture for light scalars.
Hence, predictions are made in the 2-quark model for the scalar
mesons.  The short-distance approach suffices to explain the
observed large rates of $f_0(980)K^-$ and $f_0(980)\ov K^0$ that
receive major penguin contributions from the $b\to ss\bar s$
process. When $f_0(980)$ is assigned as a four-quark bound state,
there exist extra diagrams contributing to $B\to f_0(980)K$.
Therefore, {\it a priori} the $f_0(980)K$ rate is not necessarily
suppressed for a four-quark state $f_0(980)$. The predicted $\ov
B^0\to a_0^\pm(980)\pi^\mp$ and $a_0^+(980)K^-$ rates exceed the
current experimental limits, favoring a four-quark nature for
$a_0(980)$. The penguin-dominated modes $a_0(980)K$ and
$a_0(1450)K$ receive predominant weak annihilation contributions.
There exists a two-fold experimental ambiguity in extracting the
branching ratio of $B^-\to \ov K_0^{*0}(1430)\pi^-$, which can be
resolved by measuring other $K_0^*(1430)\pi$ modes in conjunction
with the isospin symmetry consideration. Large weak annihilation
contributions are needed to explain the $K^*_0(1430)\pi$ data. The
decay $\ov B^0\to \kappa^+ K^-$ provides a nice ground for testing
the 4-quark and 2-quark nature of the $\kappa$ meson. It can
proceed through $W$-exchange and hence is quite suppressed if
$\kappa$ is made of two quarks, while it receives a tree
contribution if $\kappa$ is predominately a four-quark state.
Hence, an observation of this channel at the level of $\gsim
10^{-7}$ may imply a four-quark assignment for the $\kappa$.
Mixing-induced \CP asymmetries in penguin-dominated modes are
studied and their deviations from $\sin2\beta$ are found to be
tiny.

\pagebreak

\section{Introduction}

The first charmless $B$ decay into a scalar meson that has been
observed is $B\to f_0(980)K$. It was first measured by Belle in
the charged $B$ decays to $K^\pm\pi^\mp\pi^\pm$ and a large
branching fraction product for the $f_0(980)K^\pm$ final states
was found \cite{Bellef01} (updated in \cite{BelleKpipi} and
\cite{Bellepwave3}) and subsequently confirmed by BaBar
\cite{BaBarf0}. Recently, BaBar has searched for the decays $B\to
a_0\pi$ and $B\to a_0K$ for both charged and neutral $a_0$ mesons
\cite{BaBara0}. Many measurements of $B$ decays to other $p$-wave
mesons such as $K_0^*(1430)$, $f_0(1370)$, $f_0(1500)$,
$a_1(1260)$, $f_2(1270)$, $a_2(1320)$ and $K_2^*(1430)$ have also
been reported recently by both BaBar
\cite{BaBarpwave,BaBarKpipi,BaBarKpipi0,BaBar3pi} and Belle
\cite{Bellepwave1,Bellepwave2,Bellepwave3,BelleKpipi}. The
experimental results for the product of the branching ratios
$\B(B\to SP)$ and $\B(S\to P_1P_2)$ are summarized in Table
\ref{tab:exptBRprod}, where $S$ and $P$ stand for scalar and
pseudoscalar mesons, respectively.

These measurements should provide information on the nature of the
even-parity mesons. It is known that the identification of scalar
mesons is difficult experimentally and the underlying structure of
scalar mesons is not well established theoretically (for a review,
see e.g. \cite{Spanier,Godfrey,Close}). Studies of the mass
spectrum of scalar mesons and their strong as well as
electromagnetic decays suggest that the light scalars below or
near 1 GeV form an SU(3) flavor nonet and are predominately the
$q^2\bar q^2$ states as originally advocated by Jaffe
\cite{Jaffe}, while the scalar mesons above 1 GeV can be described
as a $q\bar q$ nonet with a possible mixing with $0^+$ $q\bar q$
and glueball states. It is hoped that through the study of $B\to
SP$, old puzzles related to the internal structure and related
parameters, e.g. the masses and widths, of light scalar mesons can
receive new understanding. For example, it has been argued that a
best candidate to distinguish the nature of the $a_0(980)$ scalar
is $\B(B^-\to a_0^-\pi^0)$ since the prediction for a four-quark
model is one order of magnitude smaller than for the two quark
assignment \cite{Delepine}.

One of the salient features of the scalar meson is that its decay
constant is either zero or small of order $m_d-m_u$,
$m_s-m_{d,u}$. Therefore, when one of the pseudoscalar mesons in
$B\to PP$ decays is replaced by the corresponding scalar, the
resulting decay pattern could be very different. Consider the
decays $B\to a_0(980)\pi$ as an example. It is expected that
$\Gamma(B^+\to a_0^+\pi^0)\ll \Gamma(B^+\to a_0^0\pi^+)$ and
$\Gamma(B^0\to a_0^+\pi^-)\ll \Gamma(B^0\to a_0^-\pi^+)$ as the
factorizable contribution proportional to the decay constant of
the scalar meson is suppressed relative to the one proportional to
the pseudoscalar meson decay constant. This feature can be checked
experimentally.

Experimentally, BaBar \cite{BaBarKpipi} and Belle
\cite{BelleKpipi} have adopted different approaches for
parametrizing the non-resonant amplitudes in the 3-body decays
$B^+\to K^+\pi^+\pi^-$. Belle found two solutions with
significantly different fractions of the $B^+\to
K_0^*(1430)^0\pi^+$ channel from the fit to $K^+\pi^+\pi^-$
events. At first sight, it appears that the solution with the
larger branching ratio, namely, $\B(B^+\to K_0^*(1430)^0\pi^+)\sim
45\times 10^{-6}$ [see Eq. (\ref{eq:K*0pi+}) below], is preferable
as it is consistent with the BaBar measurement  and supported by a
phenomenological estimate in \cite{Chernyak}. However, since the
counterpart of this decay in the 2 pseudoscalar production,
namely, $B^+\to K^0\pi^+$ has a branching ratio of order $24\times
10^{-6}$ \cite{HFAG}, one may wonder why the $K_0^{*0}\pi^+$
production is much more favorable than $K^0\pi^+$, while the
$K^{*0}_0\pi^0$ mode is comparable to $K^0\pi^0$ (see Table
\ref{tab:exptBR}). In this work, we shall examine the $K^*_0\pi$
modes carefully within the framework of QCD factorization
\cite{BBNS}.

Direct {\it CP} asymmetries in $f_0K$ and $K_0^*(1430)\pi$ modes
have been measured recently by BaBar and Belle (see Table III).
Since direct \CP violation is sensitive to the strong phases
involved in the decay processes, the comparison between theory and
experiment will provide information on the strong phases necessary
for producing the measured direct {\it CP} asymmetries.

The layout of the present paper is as follows. In Sec. II, we
extract the absolute branching ratios of $B\to SP$ from the
measured product of the branching ratios $\B(B\to SP)$ and
$\B(S\to P_1P_2)$. The physical properties of the scalar mesons
such as the quark contents, decay constants, form factors and
their light-cone distribution amplitudes are discussed in Sec.
III. We then apply QCD factorization in Sec. IV to calculate the
branching ratios and \CP asymmetries for $B\to SP$ decays. Sec. V
contains our conclusions. The factorizable amplitudes of various
$B\to SP$ decays are summarized in Appendix A. Based on the QCD
sum rule method, the decay constants and the leading twist
light-cone distribution amplitudes of the scalar mesons are
evaluated in Appendices B and C respectively.

\section{Experimental Status}

\begin{table}[t]
\caption{Experimental branching ratio products (in units of
$10^{-6}$) of $B$ decays to final states containing scalar mesons.
The third error whenever occurred represents the model
dependence.} \label{tab:exptBRprod}
\begin{ruledtabular}
\begin{tabular}{l c c  c}
Mode & BaBar
\cite{BaBarf0,BaBara0,BaBarpwave,BaBarKpipi,BaBarKpipi0, BaBar3pi}
&
Belle \cite{Bellepwave1,Bellepwave2,Bellepwave3,BelleKpipi,BelleKs2pi} & Average  \\
\hline
 $\B(B^+\to \sigma\pi^+)$ & $<4.1$ & $$ & $<4.1$ \\
 $\B(B^+\to f_0(980)K^+)\B(f_0(980)\to\pi^+\pi^-)$ &
 $9.3\pm1.0\pm0.5^{+0.3}_{-0.7}$~\footnotemark[1] & $8.8\pm0.8\pm0.7^{+0.6}_{-1.6}$ \footnotemark[1]
 &  $9.1^{+0.8}_{-1.1}$ \\
 $\B(B^+\to f_0(980)K^+)\B(f_0(980)\to K^+K^-)$ &
 & $<2.9$ & $<2.9$ \\
 $\B(B^0\to f_0(980)K^0)\B(f_0(980)\to\pi^+\pi^-)$ &
 $5.5\pm0.7\pm0.6$ & $7.60\pm1.66\pm0.59^{+0.48}_{-0.67}$ & $5.9\pm0.8$ \\
 $\B(B^+\to f_0(980)\pi^+)\B(f_0(980)\to\pi^+\pi^-)$ &
 $<3.0$ & $$ & $<3.0$ \\
 $\B(B^+\to a_0^0(980)K^+)\B(a_0(980)^0\to\eta\pi^0)$ & $<2.5$ & &
 $<2.5$ \\
 $\B(B^+\to a_0^+(980)K^0)\B(a_0(980)^+\to\eta\pi^+)$ & $<3.9$ & &
 $<3.9$ \\
 $\B(B^+\to a_0^0(980)\pi^+)\B(a_0(980)^0\to\eta\pi^0)$ & $<5.8$ & &
 $<5.8$ \\
 $\B(B^0\to a_0^-(980)K^+)\B(a_0(980)^-\to\eta\pi^-)$ & $<2.1$ & $<1.6$ &
 $<1.6$ \\
 $\B(B^0\to a_0^0(980)K^0)\B(a_0(980)^0\to\eta\pi^0)$ & $<7.8$ & &
 $<7.8$ \\
 $\B(B^0\to a_0^\mp(980)\pi^\pm)\B(a_0(980)^\mp\to\eta\pi^\mp)$ & $<5.1$ & $<2.8$ &
 $<2.8$ \\
 $\B(B^+\to f_0(1370)K^+)\B(f_0(1370)\to\pi^+\pi^-)$ & $<10.7$ & $$ &
 $<10.7$ \\
 $\B(B^+\to f_0(1370)\pi^+)\B(f_0(1370)\to\pi^+\pi^-)$ & $<3.0$ & $$ &
 $<3.0$ \\
  $\B(B^+\to f_0(1500)K^+)\B(f_0(1500)\to\pi^+\pi^-)$ & $<4.4$ & $$ &
 $<4.4$ \\
 $\B(B^+\to K^{*0}_0(1430)\pi^+)\B(K^{*0}_0(1430)\to K^+\pi^-)$ &
 $34.4\pm1.7\pm1.8^{+0.1}_{-1.4}$\footnotemark[2] & $27.9\pm1.8\pm2.6^{+8.5}_{-5.4}$\footnotemark[3] &  $$ \\
 & & $5.1\pm1.4\pm0.5^{+1.9}_{-0.5}$\footnotemark[3] & \\
 $\B(B^0\to K^{*+}_0(1430)\pi^-)\B(K^{*+}_0(1430)\to K^0\pi^+)$ &
 $$ & $30.8\pm2.4\pm2.4^{+0.8}_{-3.0}$ & $30.8^{+3.5}_{-4.5}$   \\
 $\B(B^0\to K^{*+}_0(1430)\pi^-)\B(K^{*+}_0(1430)\to K^+\pi^0)$ &
 $11.2\pm1.5\pm3.5$ & $$ & $11.2\pm3.8$\footnotemark[4] \\
 $\B(B^0\to K^{*0}_0(1430)\pi^0)\B(K^{*0}_0(1430)\to K^+\pi^-)$ &
 $7.9\pm1.5\pm2.7$ & $$ & $7.9\pm3.1$\footnotemark[4] \\
\end{tabular}
\end{ruledtabular}
\footnotetext[1]{The previously published results are
$(9.2\pm1.2^{+2.1}_{-2.6})\times 10^{-6}$ by BaBar \cite{BaBarf0}
and $(7.6\pm1.2^{+1.6}_{-1.2})\times 10^{-6}$ by Belle
\cite{BelleKpipi}.}
 \footnotetext[2]{The BaBar result is for $B^+\to (K\pi)_0^{*0}\pi^+$
 followed by $(K\pi)_0^{*0}\to K^+\pi^-$ \cite{BaBarKpipi}. The $(K\pi)_0^{*0}$ component
 consists of a nonresonant effective range term plus the
 $K^{*0}_0(1430)$ resonance itself. Using the knowledge of the
 composition of the $K^{*0}_0(1430)$ component, BaBar obtained the
 branching ratio of $B^+\to K_0^{*0}(1430)\pi^+$ as shown in Eq.
 (\ref{eq:K*0pi+}). }
 \footnotetext[3]{Two solutions
with significantly different branching ratios of the $B^+\to
K_0^*(1430)^0\pi^+$ channel but similar likelihood values were
obtained by Belle from the fit to $K^+\pi^+\pi^-$ events
\cite{BelleKpipi}. A new Belle measurement of $K^+\pi^+\pi^-$
yields $32.0\pm1.0\pm2.4^{+1.1}_{-1.9}$ for the larger solution
\cite{Bellepwave3}.}
 \footnotetext[4]{The results $\B(B^0\to K^{*+}_0(1430)\pi^-)\B
(K^{*+}_0(1430)\to K^+\pi^0)=(5.1\pm1.5^{+0.6}_{-0.7})\times
10^{-6}$ and $\B(B^0\to K^{*0}_0(1430)\pi^0)\B(K^{*0}_0(1430)\to
K^+\pi^-)=(6.1^{+1.6+0.5}_{-1.5-0.6})\times 10^{-6}$ are quoted in
\cite{BaBarKpipi0} as Belle measurements. But they will not be
included for the average as we cannot find these results in any
Belle publications.}
\end{table}

The experimental results for the product of the branching ratios
$\B(B\to SP)$ and $\B(S\to P_1P_2)$ are summarized in Table
\ref{tab:exptBRprod}. Here we shall try to determine $\B(B\to SP)$
given the information on $\B(S\to P_1P_2)$. The absolute branching
ratios for $B\to f_0(980)K$ and $f_0(980)\pi$ depend critically on
the branching fraction of $f_0(980)\to \pi\pi$. For this purpose,
we shall use the results from the most recent analysis of
\cite{Anisovich-f0}, namely, $\Gamma_{\pi\pi}=64\pm8$ MeV,
$\Gamma_{K\bar K}=12\pm1$ MeV and $\Gamma_{\rm tot}=80\pm10$ MeV
for $f_0(980)$. Therefore,
 \be
 \B(f_0(980)\to\pi^+\pi^-)=0.53\pm0.09\,, \qquad\qquad
 \B(f_0(980)\to K^+K^-)=0.08\pm0.01\,.
 \en
The obtained ratio
$r\equiv\B(f_0(980)\to\pi^+\pi^-)/\B(f_0(980)\to K^+K^-)\simeq
7.1$ is consistent with the result of $r>3.0^{+0.4}_{-0.7}$
inferred from the Belle measurements of $\B(B^+\to f_0(980)K^+\to
\pi^+\pi^-K^+)$ and $\B(B^+\to f_0(980)K^+\to K^+K^-K^+)$ (see
Table \ref{tab:exptBRprod}).

For $a_0$, we apply the Particle Data Group (PDG) average
$\Gamma(a_0\to K\ov K)/\Gamma(a_0\to \pi\eta)=0.183\pm 0.024$
\cite{PDG} to obtain
 \be \label{aKK}
 \B(a_0(980)\to \eta\pi)=0.845\pm0.017\,.
 \en
Needless to say, it is of great importance to have more precise
measurements of the branching fractions of $f_0$ and $a_0$. For
$K_0^*(1430)$ we have \cite{PDG}
 \be \label{eq:K*toKpi}
\B(K_0^{*0}\to K^+\pi^-)={2\over 3}(0.93\pm0.10), \qquad
\B(K_0^{*0}\to
 K^+\pi^0)={1\over 3}(0.93\pm0.10).
 \en

As noted in Table \ref{tab:exptBRprod}, Belle found two solutions
for the branching ratios of $B^+\to K_0^*(1430)^0\pi^+$ from the
fit to $B^+\to K^+\pi^+\pi^-$ events \cite{BelleKpipi}. BaBar
\cite{BaBarKpipi} adopted a different approach to analyze the
$K^+\pi^+\pi^-$ data by parametrizing $K^*_0(1430)^0\pi^+$ and the
non-resonant component by a single amplitude suggested by the LASS
collaboration to describe the scalar amplitude in elastic $K\pi$
scattering. As commented in \cite{BelleKpipi}, while this approach
is experimentally motivated, the use of the LASS parametrization
is limited to the elastic region of $M(K\pi)\lsim 2.0$ GeV, and an
additional amplitude is still required for a satisfactory
description of the data. Therefore, additional external
information is needed in order to resolve the ambiguity in regard
to the branching fraction of $B^+\to K_0^*(1430)^0\pi^+$:
 \be \label{eq:K*0pi+}
 \B(B^+\to K_0^*(1430)^0\pi^+)=\cases{ (37.0\pm1.8\pm1.9^{+0.1}_{-1.5}\pm4.1)\times 10^{-6}; & BaBar
 \cite{BaBarKpipi},
 \cr   (45.0\pm2.9\pm4.2^{+13.7}_{-~8.7}\pm4.8)\times 10^{-6}; & Belle~(solution I) \cite{BelleKpipi}, \cr
 (8.2\pm2.2\pm0.8^{+3.1}_{-0.8}\pm0.9)\times 10^{-6}; & Belle (solution II) \cite{BelleKpipi}, \cr}
 \en
where the fourth error is due to the uncertainty on the branching
fraction of $K^*_0(1430)\to K\pi$ [see Eq. (\ref{eq:K*toKpi})].
For the BaBar result, the uncertainty on the proportion of the
$(K\pi)^{*0}_0$ component due to the $K^{*0}_0(1430)$ resonance is
also included in the fourth error.

As shown in Sec. IV.B.3, the aforementioned ambiguity can be
resolved by measuring other $K^*_0(1430)\pi$ modes. The $\Delta
I=0$ penguin dominance implies, for example, the isospin relation
$\Gamma(B^+\to K^{*0}_0\pi^+)=\Gamma(B^0\to K^{*+}_0\pi^-)$. The
recent measurements of the three-body decays $B^0\to
K^+\pi^+\pi^0$ by BaBar  \cite{BaBarKpipi0} and $B^0\to
K_S^0\pi^+\pi^-$ by Belle \cite{BelleKs2pi} yield
 \be \label{eq:K*+pi-}
 \B(B^0\to K_0^*(1430)^+\pi^-)=\cases{ (36.1\pm4.8\pm11.3\pm3.9)\times 10^{-6}; & BaBar
 \cite{BaBarKpipi0},
 \cr   (49.7\pm3.8\pm3.8^{+1.2}_{-4.8})\times 10^{-6}; & Belle \cite{BelleKs2pi}.  \cr}
 \en
It is clear that the isospin relation is well respected by both
BaBar and Belle measurements of $K^{*0}_0\pi^+$ and
$K^{*+}_0\pi^-$ and that the smaller of the two solutions found by
Belle (solution II) is ruled out.

Experimental measurements of direct \CP asymmetries for various
$B\to SP$ decays are shown in Table \ref{tab:CP}. We see that
BaBar and Belle results for direct \CP violation are consistent
with zero.

\begin{table}[t]
\caption{Experimental branching ratios (in units of $10^{-6}$) of
$B$ decays to final states containing scalar mesons.}
\label{tab:exptBR}
\begin{ruledtabular}
\begin{tabular}{l c | l  c}
Mode & Br &  Mode & Br  \\  \hline
 $\B(B^+\to\sigma\pi^+)$ & $<4.1$\qquad &  $\B(B^0\to f_0(980)K^0)$ & $11.1\pm2.4$ \\
 $\B(B^+\to f_0(980)\pi^+)$ &  $<5.7$ &  $\B(B^0\to a_0^\mp(980)\pi^\pm)$ & $<3.3$ \footnotemark[1]  \\
 $\B(B^+\to f_0(980)K^+)$ & $17.1^{+3.3}_{-3.5}$ &  $\B(B^0\to a_0^0(980)K^0)$ & $<9.2$  \\
 $\B(B^+\to a_0^0(980)\pi^+)$ & $<6.9$ & $\B(B^0\to a_0^-(980)K^+)$ & $<1.9$  \\
 $\B(B^+\to a_0^0(980)K^+)$ & $<3.0$ & $\B(B^0\to K^{*0}_0(1430)\pi^0)$
 & $12.7\pm5.4$  \\
 $\B(B^+\to a_0^+(980)K^0)$ & $<4.6$ & $\B(B^0\to K^{*+}_0(1430)\pi^-)$
 & $47.2^{+5.6}_{-6.9}$  \\
  $\B(B^+\to K^{*0}_0(1430)\pi^+)$ & $38.2^{+4.6}_{-4.5}$ & \\
\end{tabular}
\end{ruledtabular}
\footnotetext[1]{Experimentally, one cannot separate $B^0$ from
$\ov B^0$ decays, though theoretical calculations indicate
$\Gamma(B^0\to a_0^+\pi^-)\ll \Gamma(B^0\to a_0^-\pi^+)$ (see
Table \ref{tab:theoryBR1}).}
\end{table}

\begin{table}[t]
\caption{Experimental results of direct \CP asymmetries in $B$
decays to final states containing scalar mesons.} \label{tab:CP}
\begin{ruledtabular}
\begin{tabular}{l c c c}
Mode & BaBar \cite{BaBarKpipi,BaBarKpipi0,BaBar3pi,BaBarf0K} & Belle \cite{Bellepwave3,BelleKpipi,Bellef0K,Bellef0Kcp}
& Average  \\
\hline
 $B^+\to f_0(980)K^+$ & $0.09\pm0.10\pm0.03^{+0.14}_{-0.10}$ & $-0.077\pm0.065\pm0.030^{+0.041}_{-0.016}$ &
 $-0.020^{+0.068}_{-0.065}$ \\
 $B^0\to f_0(980)K^0$ & $0.24\pm0.31\pm0.15$ &
 $-0.23\pm0.23\pm0.13$ & $-0.06\pm0.21$ \\
 $B^+\to f_0(980)\pi^+$ & $-0.50\pm0.54\pm0.06$ & $$ & $-0.50\pm0.54$ \\
 $B^+\to K^{*0}_0(1430)\pi^+$ & $-0.06\pm0.03^{+0.05}_{-0.06}$
 & $0.06\pm0.05^{+0.02}_{-0.32}$ & $-0.05^{+0.05}_{-0.08}$ \\
 $B^0\to K^{*+}_0(1430)\pi^-$ & $-0.07\pm0.12\pm0.08$ & &
 $-0.07\pm0.14$ \\
 $B^0\to K^{*0}_0(1430)\pi^0$ & $-0.34\pm0.15\pm0.11$ & &
 $-0.34\pm0.19$
\end{tabular}
\end{ruledtabular}
\end{table}

\section{Physical properties of scalar mesons}

It is known that the underlying structure of scalar mesons is not
well established theoretically (for a review, see e.g.
\cite{Spanier,Godfrey,Close}). It has been suggested that the
light scalars below or near 1 GeV--the isoscalars $f_0(600)$ (or
$\sigma$), $f_0(980)$, the isodoublet $K_0^*(800)$ (or $\kappa$)
and the isovector $a_0(980)$--form an SU(3) flavor nonet, while
scalar mesons above 1 GeV, namely, $f_0(1370)$, $a_0(1450)$,
$K^*_0(1430)$ and $f_0(1500)/f_0(1710)$, form another nonet. A
consistent picture \cite{Close} provided by the data  suggests
that the scalar meson states above 1 GeV can be identified as a
conventional $q\bar q$ nonet with some possible glue content,
whereas the light scalar mesons below or near 1 GeV form
predominately a $qq\bar q\bar q$ nonet \cite{Jaffe,Alford} with a
possible mixing with $0^+$ $q\bar q$ and glueball states. This is
understandable because in the $q\bar q$ quark model, the $0^+$
meson has a unit of orbital angular momentum and hence it should
have a higher mass above 1 GeV. On the contrary, four quarks
$q^2\bar q^2$ can form a $0^+$ meson without introducing a unit of
orbital angular momentum. Moreover, color and spin dependent
interactions favor a flavor nonet configuration with attraction
between the $qq$ and $\bar q\bar q$ pairs. Therefore, the $0^+$
$q^2\bar q^2$ nonet has a mass near or below 1 GeV. This
four-quark scenario explains naturally the mass degeneracy of
$f_0(980)$ and $a_0(980)$, the broader decay widths of
$\sigma(600)$ and $\kappa(800)$ than $f_0(980)$ and $a_0(980)$,
and the large coupling of $f_0(980)$ and $a_0(980)$ to $K\ov K$.
The four-quark flavor wave functions of light scalar mesons are
symbolically given by \cite{Jaffe}
 \be \label{4quarkw.f.}
 && \sigma=u\bar u d\bar d, \qquad\qquad f_0=s\bar s(u\bar u+d\bar d)/\sqrt{2},  \non \\
 && a_0^0={1\over\sqrt{2}}(u\bar u-d\bar d)s\bar s, \qquad a_0^+=u\bar ds\bar s,
 \qquad a_0^-=d\bar us\bar s, \non \\
 && \kappa^+=u\bar sd\bar d, \qquad \kappa^0=d\bar su\bar u,
 \qquad \bar \kappa^0=s\bar du\bar u, \qquad \kappa^-=s\bar ud\bar
 d.
 \en
This is supported by a lattice calculation \cite{Alford}.

While the above-mentioned four-quark assignment of light scalar
mesons is certainly plausible when the light scalar meson is
produced in low-energy reactions, one may wonder if the energetic
$f_0(980)$ produced in $B$ decays is dominated by the four-quark
configuration as it requires to pick up two energetic
quark-antiquark pairs to form a fast-moving light four-quark
scalar meson. The Fock states of $f_0(980)$ consist of $q\bar q$,
$q^2\bar q^2$, $q\bar q g,\cdots$, etc. Naively, it is expected
that the distribution amplitude of $f_0(980)$ would be smaller in
the four-quark model than in the two-quark picture.

In the naive 2-quark model, the flavor wave functions of the light
scalars read
 \be
 && \sigma={1\over \sqrt{2}}(u\bar u+d\bar d), \qquad\qquad f_0= s\bar s\,, \non \\
 && a_0^0={1\over\sqrt{2}}(u\bar u-d\bar d), \qquad a_0^+=u\bar d, \qquad
 a_0^-=d\bar u,  \\
 && \kappa^{+}=u\bar s, \qquad \kappa^{0}= d\bar s, \qquad
 \bar \kappa^{0}=s\bar d,\qquad \kappa^{-}=s\bar u, \non
 \en
where the ideal mixing for $f_0$ and $\sigma$ is assumed as
$f_0(980)$ is the heaviest and $\sigma$ is the lightest one in the
light scalar nonet. In this picture, $f_0(980)$ is purely an
$s\bar s$ state and this is supported by the data of $D_s^+\to
f_0\pi^+$ and $\phi\to f_0\gamma$ implying the copious $f_0(980)$
production via its $s\bar s$ component. However, there also exist
some experimental evidences indicating that $f_0(980)$ is not
purely an $s\bar s$ state. First, the observation of
$\Gamma(J/\psi\to f_0\omega)\approx {1\over 2}\Gamma(J/\psi\to
f_0\phi)$ \cite{PDG} clearly indicates the existence of the
non-strange and strange quark content in $f_0(980)$. Second, the
fact that $f_0(980)$ and $a_0(980)$ have similar widths and that
the $f_0$ width is dominated by $\pi\pi$ also suggests the
composition of $u\bar u$ and $d\bar d$ pairs in $f_0(980)$; that
is, $f_0(980)\to\pi\pi$ should not be OZI suppressed relative to
$a_0(980)\to\pi\eta$. Therefore, isoscalars $\sigma(600)$ and
$f_0$ must have a mixing
 \be
 |f_0(980)\ra = |s\bar s\ra\cos\theta+|n\bar n\ra\sin\theta,
 \qquad |\sigma(600)\ra = -|s\bar s\ra\sin\theta+|n\bar n\ra\cos\theta,
 \en
with $n\bar n\equiv (\bar uu+\bar dd)/\sqrt{2}$.

Experimental implications for the $f_0\!-\!\sigma$ mixing angle
have been discussed in detail in \cite{ChengDSP}:
 \be
 J/\psi\to f_0\phi,~f_0\omega [26] \quad &\Rightarrow &\quad
 \theta=(34\pm6)^\circ~~{\rm or}~~\theta=(146\pm 6)^\circ,  \non \\
 R=4.03\pm0.14~[26] \quad &\Rightarrow &\quad
 \theta=(25.1\pm0.5)^\circ~~{\rm or}~~\theta=(164.3\pm 0.2)^\circ, \non \\
 R=1.63\pm0.46~[26] \quad &\Rightarrow &\quad
 \theta=(42.3^{+8.3}_{-5.5})^\circ~~{\rm or}~~
 \theta=(158\pm 2)^\circ,  \non \\
 \phi\to f_0\gamma,~f_0\to\gamma\gamma~[27] \quad &\Rightarrow &\quad
 \theta=(5\pm5)^\circ~~{\rm or}~~\theta=(138\pm6)^\circ,  \non \\
 {\rm QCD~sum~rules~and}~f_0~{\rm data}~[28] \quad &\Rightarrow &\quad
 \theta=(27\pm13)^\circ~~{\rm or}~~\theta=(153\pm13)^\circ, \non \\
 {\rm QCD~sum~rules~and}~a_0~{\rm data}~[28] \quad &\Rightarrow &\quad
 \theta=(41\pm11)^\circ~~{\rm or}~~\theta=(139\pm11)^\circ,
 \en
where $R\equiv g^2_{f_0K^+K^-}/g^2_{f_0\pi^+\pi^-}$ measures the
ratio of the $f_0(980)$ coupling to $K^+K^-$ and $\pi^+\pi^-$. In
short, $\theta$ lies in the ranges of $25^\circ<\theta<40^\circ$
and $140^\circ<\theta< 165^\circ$. Note that the phenomenological
analysis of the radiative decays $\phi\to f_0(980)\gamma$ and
$f_0(980)\to\gamma\gamma$ favors the second solution, namely,
$\theta=(138\pm 6)^\circ$. The fact that phenomenologically there
does not exist a unique mixing angle solution may already indicate
that $f_0(980)$ and $\sigma$ are not purely $q\bar q$ bound
states.

Likewise, in the four-quark scenario for light scalar mesons, one
can also define a similar $f_0-\sigma$ mixing angle
  \be
 |f_0(980)\ra =|n\bar ns\bar s\ra\cos\phi
 +|u\bar u d\bar d\ra\sin\phi, \qquad |\sigma(600)\ra = -|n\bar ns
 \bar s\ra\sin\phi+|u\bar u d\bar d\ra\cos\phi.
 \en
It has been shown that $\phi=174.6^\circ$ \cite{Maiani}.

\subsection{Decay constants} To proceed we first discuss the decay
constants of the pseudoscalar meson $P$ and the scalar meson $S$
defined by
 \be \label{eq:decayc}
 && \la P(p)|\bar q_2\gamma_\mu\gamma_5 q_1|0\ra=-if_Pp_\mu, \qquad
 \la S(p)|\bar q_2\gamma_\mu q_1|0\ra=f_S p_\mu,
 \qquad \la S|\bar q_2q_1|0\ra=m_S\bar f_S.
 \en
If the scalar meson is a four-quark bound state, it is pertinent
to consider the interpolating current $j_S$, for example,
 \be
  j_{f_0}={1\over\sqrt{2}}\epsilon_{abc}\epsilon_{dec}[(u_a^TC\gamma_5
  s_b)(\bar u_d\gamma_5C\bar s^T_e)+(u\to d)],
 \en
with $a,b,c\cdots$ being the color indices and $C$ the charge
conjugation matrix. The coupling of the scalar meson $S$ to the
scalar current $j_S$ is parametrized in terms of the scalar decay
constant $F_S$ defined by
 \be
 \la S|j_S|0\ra=\sqrt{2}F_Sm_S^4.
 \en

The neutral scalar mesons $\sigma$, $f_0$ and $a_0^0$ cannot be
produced via the vector current owing to charge conjugation
invariance or conservation of vector current:
 \be
 f_{\sigma}=f_{f_0}=f_{a_0^0}=0.
 \en
For other scalar mesons, the vector decay constant $f_S$ and the
scale-dependent scalar decay constant $\bar f_S$ are related by
equations of motion
 \be \label{eq:EOM}
 \mu_Sf_S=\bar f_S, \qquad\quad{\rm with}~~\mu_S={m_S\over
 m_2(\mu)-m_1(\mu)},
 \en
where $m_{2}$ and $m_{1}$ are the running current quark masses.
Therefore, contrary to the case of pseudoscalar mesons, the vector
decay constant of the scalar meson, namely, $f_S$, vanishes in the
SU(3) or isospin limit. For example, the vector decay constant of
$K^{*+}_0$ ($a^+_0$) is proportional to the mass difference
between the constituent $s$ ($d$) and $u$ quarks; that is, the
decay constants of $K^*_0(1430)$ and the charged $a_0(980)$ are
suppressed. In short, the vector decay constants of scalar mesons
are either zero or small.

For light scalar mesons, only two estimates of $F_S$ in the
four-quark scenario are available in the literature
\cite{Latorre,Brito} and all other decay constant calculations are
done in the 2-quark picture for light scalars. The results of
$F_S$ are \cite{Brito}
 \be
 F_\sigma=(7.5\pm 1.0)\,{\rm MeV}, \quad
 F_\kappa=(1.6\pm 0.3)\,{\rm MeV}, \quad
 F_{f_0}=F_{a_0}=(1.1\pm 0.1)\,{\rm MeV}.
 \en

We now turn to the model calculations in which the light scalar is
assumed to be a two-quark bound state. Based on the finite-energy
sum rule, Maltman obtained \cite{Maltman}
 \be \label{eq:Maltman}
 f_{a_0(980)}=1.1\pm0.2\,{\rm MeV},\qquad  f_{a_0(1450)}=0.7\pm0.1\,{\rm MeV},\qquad
 f_{K_0^*}=42\pm2\,{\rm MeV},
 \en
in accordance with the ranges estimated by Narison \cite{Narison}
 \be
  f_{a_0(980)}=0.7\sim2.5\,{\rm MeV},\qquad f_{K_0^*}=33\sim 46\,{\rm
  MeV}.
 \en
A different calculation of the scalar meson decay constants based
on the generalized NLJ model yields \cite{Shakin}
 \be
 f_{a_0(980)}=1.6\,{\rm MeV},\qquad  f_{a_0(1450)}=0.4\,{\rm MeV},\qquad
 f_{K_0^*}=31\,{\rm MeV}.
 \en
Note that in \cite{Maltman} and \cite{Shakin} the $a_0$ decay
constant is defined with an extra factor of $(m_s-m_u)/(m_d-m_u)$.
We have taken the quark masses $m_s=119$ MeV, $m_d=6.3$ MeV and
$m_u=3.5$ MeV at $\mu=1$ GeV  to convert it into our convention.
Based on the QCD sum rule method, a recent estimate of the
$K_0^*(1430)$ scalar decay constant yields $\bar
f_{K^*_0}=427\pm85$ MeV at $\mu\sim 1$ GeV \cite{Du} which
corresponds to $f_{K^*_0}=34\pm7$ MeV.\footnote{The estimate by
Chernyak \cite{Chernyak}, namely, $f_{K^*_0}=(70\pm10)$ MeV, seems
to be too large.}

Because of the $f_0-\sigma$ mixing, we shall treat $f_0$ and
$\sigma$ separately. Just like the case of $\eta$ and $\eta'$,
each meson is described by four decay constants:\footnote{Note
that $\la a_0|\bar ss|0\ra=0$ even when $a_0$ is a four-quark
bound state. This is because $\bar ss$ is an isospin singlet while
$a_0$ is an isospin triplet.}
 \be \label{eq:bardecayc}
 \la f_0|\bar uu|0\ra={1\over\sqrt{2}}m_{f_0}\bar f_{f_0}^n, &&
 \qquad \la f_0|\bar ss|0\ra=m_{f_0}\bar f_{f_0}^s, \non \\
 \la \sigma|\bar uu|0\ra={1\over\sqrt{2}}m_\sigma\bar f_{\sigma}^n, &&
 \qquad \la \sigma|\bar ss|0\ra=m_{\sigma}\bar f_{\sigma}^s,
 \en
or
 \be \label{eq:tildedecayc}
 \la f_0^n|\bar uu|0\ra={1\over\sqrt{2}}m_{f_0}\tilde f_{f_0}^n,
 &\qquad&
 \la f_0^s|\bar ss|0\ra=m_{f_0}\tilde f_{f_0}^s, \non \\
  \la \sigma^n|\bar uu|0\ra={1\over\sqrt{2}}m_\sigma\tilde f_\sigma^n,
 &\qquad&
 \la \sigma^s|\bar ss|0\ra=m_\sigma\tilde f_\sigma^s,
 \en
where $f_0^n,\,\sigma^n= \bar nn$ and $f_0^s,\,\sigma^s= \bar ss$.
It follows that \cite{FKS}
 \be
 \bar f_{f_0}^n=\tilde f_{f_0}^n\sin\theta, \qquad
 \bar f_{f_0}^s=\tilde f_{f_0}^s\cos\theta, \qquad
 \bar f_{\sigma}^n=\tilde f_{\sigma}^n\cos\theta, \qquad
 \bar f_{\sigma}^s=-\tilde f_{\sigma}^s\sin\theta.
 \en
Using the QCD sum-rule method, the scalar decay constant $\tilde
f^s_{f_0}$ defined in Eq. (\ref{eq:tildedecayc}) has been
estimated in \cite{Fazio} and \cite{Bediaga} with similar results,
namely, $\tilde f^s_{f_0}\approx 180$ MeV at a typical hadronic
scale. Taking into account the scale dependence of $\tilde
f_{f_0}$ and radiative corrections to the quark loops in the OPE
series,  we have made a careful evaluation of the scalar decay
constant in \cite{CYf0K} using the sum rule approach. Our updated
results for $\tilde f^s_{f_0}$ and $\tilde f_{a_0}$ of order 370
MeV at $\mu=1$ GeV (see Appendix B) are much larger than previous
estimates.\footnote{The decay constants $\tilde f_{f_0}^s$ and
$\tilde f_{f_0}^n$ have been determined separately in \cite{CYf0K}
using the sum rule approach and they are found to be very close.
Hence, for simplicity, we shall assume $\tilde f_{f_0}^s=\tilde
f_{f_0}^n$ in the present work.}
Note that taking $f_{a_0}=1.1$ MeV from Eq. (\ref{eq:Maltman})
leads to $\bar f_{a_0}(1\,{\rm GeV})\approx 385$ MeV, which is
also very similar to our estimate. Therefore, a typical scalar
decay constant of the scalar meson is above 300 MeV. In Appendix B
we give a complete summary on the sum rule estimates of scalar
meson decay constants.

\subsection{Light-Cone Distribution Amplitudes}

The twist-2 light-cone distribution amplitude (LCDA) $\Phi_S(x)$
and twist-3 $\Phi_S^s(x)$ and $\Phi_S^\sigma(x)$ for the scalar
meson $S$ made of the quarks $q_2\bar q_1$ are given by
 \be \label{eq:wfdef}
 \la S(p)|\bar q_2(z_2)\gamma_\mu q_1(z_1)|0\ra &=& p_\mu \int
 ^1_0 dx e^{i(xp\cdot z_2+\bar xp\cdot z_1)}\Phi_S(x), \non \\
 \la S(p)|\bar q_2(z_2) q_1(z_1)|0\ra &=& m_S\int
 ^1_0 dx e^{i(xp\cdot z_2+\bar xp\cdot z_1)}\Phi_S^s(x), \non \\
 \la S(p)|\bar q_2(z_2)\sigma_{\mu\nu}q_1(z_1)|0\ra &=& -m_S(p_\mu
 z_\nu-p_\nu z_\mu) \int
 ^1_0 dx e^{i(xp\cdot z_2+\bar xp\cdot z_1)}{\Phi_S^\sigma(x)\over 6},
 \en
with $z=z_2-z_1$, $\bar x=1-x$, and their normalizations are
 \be \label{eq:wfnor}
 \int_0^1dx \Phi_S(x)=f_S, \qquad \int_0^1dx \Phi_S^s(x)=
 \int_0^1dx \Phi_S^\sigma(x)=\bar f_S.
 \en
The definitions of LCDAs given in Eq. (\ref{eq:wfdef}) can be
combined into a single matrix element
 \be \label{eq:generalwf}
 && \la S(p)|\bar q_{2\beta}(z_2)q_{1\alpha}(z_1)|0\ra \non  \\
 &&= {1\over
 4}\int_0^1 dxe^{i(xp\cdot z_2+\bar xp\cdot z_1)}\Bigg\{
 p\!\!\!/\Phi_S(x)+m_S\left(\Phi_S^s(x)-\sigma_{\mu\nu}p^\mu
 z^\nu{\Phi_S^\sigma(x)\over 6}\right)\Bigg\}_{\alpha\beta}.
 \en
In general, the twist-2 light-cone distribution amplitude $\Phi_S$
has the form
 \be  \label{eq:Swf}
 \Phi_S(x,\mu)=\bar f_S(\mu)\,6x(1-x)\left[B_0(\mu)+\sum_{m=1}^\infty
 B_m(\mu)\,C_m^{3/2}(2x-1)\right],
 \en
where $B_m$ are Gegenbauer moments and $C_m^{3/2}$ are the
Gegenbauer polynomials. The normalization condition
(\ref{eq:wfnor}) indicates
 \be
 B_0=\mu_S^{-1},
 \en
where we have applied Eq. (\ref{eq:EOM}) and neglected the
contributions from the even Gegenbaur moments. It is clear that
the $B_0$ term is either zero or small of order $m_d-m_u$ or
$m_s-m_{d,u}$, so are other even Gegenbaur moments [see also Eq.
(\ref{eq:Bm})]. For the neutral scalar mesons $f_0$, $a_0^0$ and
$\sigma$, $B_0=0$ and only odd Gegenbauer polynomials contribute.
The LCDA also can be recast to the form
 \be \label{eq:twist2wf}
 \Phi_S(x,\mu)=f_S\,6x(1-x)\left[1+\mu_S\sum_{m=1}^\infty
 B_m(\mu)\,C_m^{3/2}(2x-1)\right],
 \en
which we shall use for later purposes. Since $\mu_S\gg 1$ and even
Gegenbauer coefficients are suppressed, it is clear that the LCDA
of the scalar meson is dominated by the odd Gegenabuer moments. In
contrast, the odd Gegenbauer moments vanish for the $\pi$ and
$\rho$ mesons.

When the three-particle contributions are neglected, the twist-3
two-particle distribution amplitudes are determined by the
equations of motion, leading to
 \be
 (1-2x)\Phi_S^s(x)={\left(\Phi_S^\sigma(x)\right)'\over 6},
 \en
where use of Eq. (\ref{eq:wfdef}) has been made. This means that
we shall take the asymptotic forms
 \be \label{eq:twist3wf}
 \Phi^s_S(x)=\bar f_S , \qquad \Phi^\sigma_S(x)=\bar f_S\, 6x(1-x),
 \en
recalling that it has been shown to the leading conformal
expansion, the asymptotic forms of the twist-3 distribution
amplitudes are the same as that for the pseudoscalar mesons
\cite{Braun}. The corresponding light-cone projection operator of
Eq. (\ref{eq:generalwf}) in momentum space can be obtained by
assigning momenta \cite{BBNS}
 \be
 k_1^\mu=xp^\mu+k_\bot^\mu+{\vec{k}^2_\bot\over 2x\,p\cdot\bar p}\bar
 p^\mu, \qquad  k_2^\mu=\bar xp^\mu-k_\bot^\mu+{\vec{k}^2_\bot\over 2\bar x\,p\cdot\bar p}\bar
 p^\mu,
 \en
to the quark and antiquark  in the scalar meson, where $\bar p$ is
a light-like vector whose 3-components point into the opposite
direction of $\vec{p}$. As stressed in \cite{BBNS}, the collinear
approximation for the parton momentum (e.g. $k_1=xp$ and $k_2=\bar
xp$) can be taken only after the light-cone projection has been
applied. The light-cone projection operator of the scalar meson in
momentum space then reads
 \be
 M^S_{\alpha\beta}= {1\over 4}\Bigg(
 p\!\!\!/\Phi_S(x)+m_S{ k\!\!\!/_2 k\!\!\!/_1\over k_2
 \cdot k_1}\Phi_S^s(x)\Bigg)_{\alpha\beta},
 \en
where use of Eq. (\ref{eq:twist3wf}) has been made. By comparison,
the longitudinal part of the projection operator for the vector
meson is given by \cite{BN}
 \be
 (M^V_\|)_{\alpha\beta}=-{if_V\over
 4}\left(p\!\!\!/\Phi_V(x)-{m_Vf_V^\bot\over f_V}\,{ k\!\!\!/_2 k\!\!\!/_1\over k_2
 \cdot k_1}\Phi_v(x)\right)_{\alpha\beta},
 \en
where the definitions for the twist-3 function $\Phi_v(x)$ and the
transverse decay constant $f_V^\bot$ can be found in \cite{BN}.
Therefore, the hard-scattering kernels for $SP$  mesons in the
final state can be obtained from those for $VP$ by performing the
replacements $f_V\Phi_V(x)\to i\Phi_S(x)$ and
$m_Vf_V^\bot\Phi_v(x)\to -im_S\Phi_S^s(x)$, recalling that the
normalization for $\Phi_V$ and $\Phi_v$ is given by \cite{BN}
 \be
 \int_0^1dx\Phi_V(x)=1, \qquad  \int^1_0 dx\Phi_v(x)=0.
 \en

Just as the decay constants for $f_0(980)$ and $\sigma$, their
LCDAs should also be treated separately. The twist-2 and twist-3
distribution amplitudes $\Phi_S^{(q)}$ and $\Phi_S^{(q)s}$
$(q=n,s)$ \footnote{The quark flavor $s$ should not be confused
with the superscript $s$ for the twist-3 LCDA $\Phi^s(x)$.},
respectively, are given by
 \be
 \la S^{(n)}(p)|\bar n(z)\gamma_\mu n(0)|0\ra &=& p_\mu\int
 ^1_0 dx e^{ixp\cdot z}\Phi_S^{(n)}(x), \non \\
 \la S^{(s)}(p)|\bar s(z)\gamma_\mu s(0)|0\ra &=& p_\mu\int
 ^1_0 dx e^{ixp\cdot z}\Phi_S^{(s)}(x), \non \\
 \la S^{(n)}(p)|\bar n(z) n(0)|0\ra &=& m_S^{(n)}\int
 ^1_0 dx e^{ixp\cdot z}\Phi_S^{(n)s}(x), \non \\
 \la S^{(s)}(p)|\bar s(z) s(0)|0\ra &=& m_S^{(s)}\int
 ^1_0 dx e^{ixp\cdot z}\Phi_S^{(s)s}(x).
 \en
They satisfy the relations $\Phi_S(x)=-\Phi_S(1-x)$ due to charge
conjugation invariance (that is, the distribution amplitude
vanishes at $x=1/2$) and $\Phi_S^s(x)=\Phi_S^s(1-x)$ so that
 \be
\int^1_0dx\,\Phi_S^{(n,s)}(x)=0, \qquad
\int^1_0dx\,\Phi_S^{(n,s)s}(x)=\tilde f_S^{n,s},
 \en
with $\tilde f_S^{n,s}$ being defined in Eq.
(\ref{eq:tildedecayc}). Hence, the light-cone distribution
amplitudes for $S=f_0,\sigma$ read
 \be \label{eq:DAf0}
 \Phi_S^{(n,s)}(x,\mu)=\tilde f_S^{n,s}\,6x(1-x)\sum_{m=1,3,5,\cdots}B_m^{(n,s)}
 (\mu)C_m^{3/2}(2x-1).
 \en
The LCDAs are
 \be
 \Phi_{f_0}(x,\mu)=\Phi_{f_0}^{(s)}\cos\theta+\Phi_{f_0}^{(n)}\sin\theta,
 \qquad
 \Phi_\sigma(x,\mu)=-\Phi_\sigma^{(s)}\sin\theta+\Phi_\sigma^{(n)}\cos\theta.
 \en
Since the $B_0$ term in the LCDA for the charged $a_0$ is of order
$m_d-m_u$, it can be safely neglected. Hence, in practice we shall
use the same LCDA for both neutral and charged $a_0$ scalar
mesons.

Based on the QCD sum rule technique, the Gegenbauer moments in Eq.
(\ref{eq:DAf0}) have been evaluated in  \cite{CYf0K} up to $m=5$.
For an updated analysis, see Appendix C. Note that our result
$\bar f_{a_0}B_1^{a_0}=-340$ MeV is much larger than the estimate
of $|\bar fB_1|_{_{a_0}}\approx 100$ MeV at $\mu=m_b$ inferred
from the analysis in \cite{Diehl} (see Eq. (52) of \cite{Diehl}).

For pseudoscalar mesons,  the asymptotic forms for twist-2 and
twist-3 distribution amplitudes for pseuodscalar mesons are
 \be
 \Phi_P(x)=f_P6x(1-x), \qquad \Phi_P^p(x)=f_P, \qquad
 \Phi_P^\sigma(x)=f_P6x(1-x).
 \en

\subsection{Form factors}

Form factors for $B\to P,S$ transitions are defined by \cite{BSW}
 \be \label{m.e.}
 \la P(p')|V_\mu|B(p)\ra &=& \left(P_\mu-{m_B^2-m_P^2\over q^2}\,q_ \mu\right)
F_1^{BP}(q^2)+{m_B^2-m_P^2\over q^2}q_\mu\,F_0^{BP}(q^2), \non \\
\la S(p')|A_\mu|B(p)\ra &=& -i\Bigg[\left(P_\mu-{m_B^2-m_S^2\over
q^2}\,q_ \mu\right) F_1^{BS}(q^2)   +{m_B^2-m_S^2\over
q^2}q_\mu\,F_0^{BS}(q^2)\Bigg],
 \en
where $P_\mu=(p+p')_\mu$, $q_\mu=(p-p')_\mu$.  As shown in
\cite{CCH}, a factor of $(-i)$ is needed in $B\to S$ transition in
order for the $B\to S$ form factors to be positive. This also can
be checked from heavy quark symmetry \cite{CCH}.

Various form factors for $B\to S$ transitions have been evaluated
in the relativistic covariant light-front quark model \cite{CCH}.
In this model form factors are first calculated in the spacelike
region and their momentum dependence is fitted to a 3-parameter
form
  \be \label{eq:FFpara}
 F(q^2)=\,{F(0)\over 1-a(q^2/m_{B}^2)+b(q^2/m_{B}^2)^2}\,.
 \en
The parameters $a$, $b$ and $F(0)$ are first determined in the
spacelike region. This parametrization is then analytically
continued to the timelike region to determine the physical form
factors at $q^2\geq 0$.  The results relevant for our purposes are
summarized in Table \ref{tab:FF}. Note that the calculation of $B$
to scalar meson form factors in \cite{CCH} coauthored by two of us
is for the case where the scalar meson is made of $q\bar q'$
quarks. Since it is possible that $K_0^*(1430), a_0(1450),
f_0(1500)$ are the first excited states of $\kappa, a_0(980)$ and
$f_0(980)$, respectively, we also extend the calculation to the
case where $K_0^*(1430)$ and $a_0(1450)$ are first excited states
by working out their wave functions from a
simple-harmonic-oscillator-type potential. The resultant form
factors are shown in Table \ref{tab:FF}.


\begin{table}[t]
\caption{Form factors of $B\to\pi,K,a_0(1450),K_0^*(1430)$
transitions obtained in the covariant light-front model
\cite{CCH}. } \label{tab:FF}
\begin{ruledtabular}
\begin{tabular}{| l c c c c || l c c c c |}
~~~$F$~~~~~
    & $F(0)$~~~~~
    & $F(q^2_{\rm max})$~~~~
    &$a$~~~~~
    & $b$~~~~~~
& ~~~ $F$~~~~~
    & $F(0)$~~~~~
    & $F(q^2_{\rm max})$~~~~~
    & $a$~~~~~
    & $b$~~~~~~
 \\
    \hline
$F^{B\pi}_1$
    & $0.25$
    & $1.16$
    & 1.73
    & 0.95
& $F^{B\pi}_0$
    & 0.25
    & 0.86
    & 0.84
    & $0.10$
    \\
$F^{BK}_1$
    & $0.35$
    & $2.17$
    & 1.58
    & 0.68
& $F^{BK}_0$
    & 0.35
    & 0.80
    & 0.71
    & $0.04$
    \\
$F^{Ba_0(1450)}_1$
    & $0.26$
    & $0.68$
    & 1.57
    & 0.70
 &$F^{Ba_0(1450)}_0$
    & 0.26
    & 0.35
    & $0.55$
    & 0.03 \\
    & $0.21$\footnotemark[1]
    & $0.52$\footnotemark[1]
    & 1.66\footnotemark[1]
    & 1.00\footnotemark[1]
 &
    & 0.21\footnotemark[1]
    & 0.33\footnotemark[1]
    & $0.73$\footnotemark[1]
    & 0.09\footnotemark[1]
    \\
$F^{BK^*_0}_1$
    & $0.26$
    & $0.70$
    & 1.52
    & 0.64
&$F^{BK^*_0}_0$
    & 0.26
    & 0.33
    & 0.44
    & 0.05
    \\
    & $0.21$\footnotemark[1]
    & $0.52$\footnotemark[1]
    & 1.59\footnotemark[1]
    & 0.91\footnotemark[1]
&
    & 0.21\footnotemark[1]
    & 0.30\footnotemark[1]
    & 0.59\footnotemark[1]
    & 0.09\footnotemark[1]
    \\
\end{tabular}
\end{ruledtabular}
\footnotetext[1]{Form factors obtained by considering the scalar
meson above 1 GeV as the first excited state of the corresponding
light scalar meson.}
\end{table}


Assuming that the light scalar mesons are the bound states of
$q\bar q$, form factors for $B$ to light scalar mesons also can be
estimated in this approach. Taking the decay constants of
$f_0(980)$ and $a_0(980)$ estimated in Appendix B, it is found
that the form factor of $B$ to $f_0(980)$ or $a_0(980)$ is of
order 0.25 at $q^2=0$. Therefore, the form factor
$F_0^{Ba_0(980)}$ is not necessarily smaller than $F_0^{B\pi}$.
This is understandable because the $a_0(980)$ distribution
amplitude peaks at $x\sim 0.25$ and $x\sim 0.75$ while the pion
LCDA peaks at $x=1/2$. As pointed out in \cite{Diehl}, since
$\Phi_{a_0}$ is more pronounced towards the endpoints $x=0$ and
$x=1$, it can have a greater overlap with the highly asymmetric
wave function of the $B$ meson than the pion wave function can.
Consequently, the $B$ to $a_0(980)$ transition form factor is
anticipated to be at least of the same order as the $B\to\pi$
case. Note that based on the light-cone sum rules, Chernyak
\cite{Chernyak} has estimated the $B\to a_0(1450)$ transition form
factor and obtained $F_{0}^{Ba_0(1450)}(0)=0.46$, while our result
is 0.26 and is similar to the $B\to\pi$ form factor at $q^2=0$. We
will make a comment on this when discussing the  decay $\ov B^0\to
a_0^+(1450)\pi^-$ in Sec. IV.B.

\section{$B\to SP$ decays}

\subsection{Decay amplitudes in QCD factorization}

 We shall use the QCD factorization approach
\cite{BBNS,BN} to study the short-distance contributions to the
decays $B\to f_0(980)K,~K_0^*(1430)\pi$, and $a_0\pi,~a_0K$ for
$a_0=a_0(980)$ and $a_0(1450)$. In QCD factorization, the
factorization amplitudes of above-mentioned decays are summarized
in Appendix A. The effective parameters $a_i^p$ with $p=u,c$ in
Eq. (\ref{eq:SDAmp}) can be calculated in the QCD factorization
approach \cite{BBNS}. They are basically the Wilson coefficients
in conjunction with short-distance nonfactorizable corrections
such as vertex corrections and hard spectator interactions. In
general, they have the expressions \cite{BBNS,BN}
 \be
 a_i^p(M_1M_2) &=& c_i+{c_{i\pm1}\over N_c}
  +{c_{i\pm1}\over N_c}\,{C_F\alpha_s\over
 4\pi}\Big[V_i(M_2)+{4\pi^2\over N_c}H_i(M_1M_2)\Big]+P_i^p(M_2),
 \en
where $i=1,\cdots,10$,  the upper (lower) signs apply when $i$ is
odd (even), $c_i$ are the Wilson coefficients,
$C_F=(N_c^2-1)/(2N_c)$ with $N_c=3$, $M_2$ is the emitted meson
and $M_1$ shares the same spectator quark with the $B$ meson. The
quantities $V_i(M_2)$ account for vertex corrections,
$H_i(M_1M_2)$ for hard spectator interactions with a hard gluon
exchange between the emitted meson and the spectator quark of the
$B$ meson and $P_i(M_2)$ for penguin contractions. The vertex and
penguin corrections for $SP$ final states have the same
expressions as those for $PP$ states and can be found in
\cite{BBNS,BN}. Using the general LCDA
 \be
 \Phi_M(x,\mu)=f_M6x(1-x)\left[1+\sum_{n=1}^\infty
 \alpha_n^M(\mu)C_n^{3/2}(2x-1)\right]
 \en
with $\alpha_n=\mu_SB_n$ for the scalar meson [see Eq.
(\ref{eq:twist2wf})]  and applying Eq. (37) in \cite{BN} for
vertex corrections, we obtain (apart from the decay constant
$f_M$)
 \be
 V_i(M) &=& 12\ln{m_b\over\mu}-18-{1\over
 2}-3i\pi+\left({11\over
 2}-3i\pi\right)\alpha_1^M-{21\over 20}\alpha_2^M+\left({79\over 36}-{2i\pi\over
 3}\right)\alpha_3^M+\cdots, \non \\
 \en
for $i=1-4,9,10$,
 \be
 V_i(M) &=& -12\ln{m_b\over\mu}+6-{1\over
 2}-3i\pi-\left({11\over
 2}-3i\pi\right)\alpha_1^M-{21\over 20}\alpha_2^M-\left({79\over 36}-{2i\pi\over
 3}\right)\alpha_3^M+\cdots, \non \\
 \en
for $i=5,7$ and $V_i(M_2)=-6$ for $i=6,8$ in the NDR scheme for
$\gamma_5$. The expressions of $V_i(M)$ up to the $\alpha_2^M$
term are the same as that in \cite{BBNS}.

As for the hard spectator function $H$, it reads
 \be \label{eq:Hi}
 H_i(M_1M_2) &=& {1\over
f_{M_2}F_0^{BM_1}(0)m^2_B}\int^1_0 {d\rho\over\rho}\,
\Phi_B(\rho)\int^1_0 {d\xi\over \bar\xi} \,\Phi_{M_2}(\xi)\int^1_0
{d\eta\over
\deta}\left[\Phi_{M_1}(\eta)+r_\chi^{M_1}\,{\bar\xi\over
\xi}\,\Phi_{m_1}(\eta)\right], \non \\
 \en
for $i=1-4,9,10$,
 \be \label{eq:Hi57}
 H_i(M_1M_2) &=& -{1\over f_{M_2}
F_0^{BM_1}(0)m^2_B}\int^1_0 {d\rho\over\rho}\,
\Phi_B(\rho)\int^1_0 {d\xi\over \xi} \,\Phi_{M_2}(\xi)\int^1_0
{d\eta\over \deta}\left[\Phi_{M_1}(\eta)+r_\chi^{M_1}\,{\xi\over
\bar\xi}\,\Phi_{m_1}(\eta)\right], \non \\
 \en
for $i=5,7$ and $H_i=0$ for $i=6,8$, where $\bar\xi\equiv 1-\xi$
and $\bar\eta\equiv 1-\eta$, $\Phi_M$ ($\Phi_m$) is the twist-2
(twist-3) light-cone distribution amplitude of the meson $M$. The
ratios $r_\chi^P$, $r_\chi^V$ and $r_\chi^S$ are defined in Eqs.
(\ref{eq:rchiP}) and (\ref{eq:rchiS}). As shown in Appendix A, the
factorizable amplitudes $A_{PS}$ and $A_{SP}$ have an opposite
relative sign [see Eq. (\ref{eq:replacementII})] and  one has to
replace $r_\chi^V$ by $-r_\chi^S$ when $M_1$ is a scalar meson.
This amounts to changing the sign of the first term in the
expression of $H_i(M_1M_2)$ for a scalar meson $M_1$.

Weak annihilation contributions are described by the terms $b_i$,
and $b_{i,{\rm EW}}$ in Eq. (\ref{eq:SDAmp}) which have the
expressions
 \be
 b_1 &=& {C_F\over N_c^2}c_1A_1^i, \qquad\quad b_3={C_F\over
 N_c^2}\left[c_3A_1^i+c_5(A_3^i+A_3^f)+N_cc_6A_3^f\right], \non \\
 b_2 &=& {C_F\over N_c^2}c_2A_1^i, \qquad\quad b_4={C_F\over
 N_c^2}\left[c_4A_1^i+c_6A_2^f\right], \non \\
 b_{\rm 3,EW} &=& {C_F\over
 N_c^2}\left[c_9A_1^{i}+c_7(A_3^{i}+A_3^{f})+N_cc_8A_3^{i}\right],
 \non \\
 b_{\rm 4,EW} &=& {C_F\over
 N_c^2}\left[c_{10}A_1^{i}+c_8A_2^{i}\right],
 \en
where the subscripts 1,2,3 of $A_n^{i,f}$ denote the annihilation
amplitudes induced from $(V-A)(V-A)$, $(V-A)(V+A)$ and
$(S-P)(S+P)$ operators, respectively, and the superscripts $i$ and
$f$ refer to gluon emission from the initial and final-state
quarks, respectively.  Their explicit expressions are given by
 \be \label{eq:ann}
 A_1^{i}&=& \int\cdots \cases{
 \left( \Phi_{M_2}(x)\Phi_{M_1}(y)\left[{1\over y(1-x\bar y)}+{1\over \bar
 x^2y}\right]-r_\chi^{M_1}r_\chi^{M_2}\Phi_{m_2}(x)\Phi_{m_1}(y)\,{2\over \bar
 xy}\right); & for~$M_1M_2=PS$,  \cr
 \left( \Phi_{M_2}(x)\Phi_{M_1}(y)\left[{1\over y(1-x\bar y)}+{1\over \bar
 x^2y}\right]+r_\chi^{M_1}r_\chi^{M_2}\Phi_{m_2}(x)\Phi_{m_1}(y)\,{2\over \bar
 xy}\right); & for~$M_1M_2=SP$, } \non  \\
 A_2^{i}&=& \int\cdots \cases{
 \left( -\Phi_{M_2}(x)\Phi_{M_1}(y)\left[{1\over \bar x(1-x\bar y)}+{1\over \bar
 xy^2}\right]+r_\chi^{M_1}r_\chi^{M_2}\Phi_{m_2}(x)\Phi_{m_1}(y)\,{2\over \bar
 xy}\right); & for~$M_1M_2=PS$,  \cr
 \left( -\Phi_{M_2}(x)\Phi_{M_1}(y)\left[{1\over \bar x(1-x\bar y)}+{1\over \bar
 xy^2}\right]-r_\chi^{M_1}r_\chi^{M_2}\Phi_{m_2}(x)\Phi_{m_1}(y)\,{2\over \bar
 xy}\right); & for~$M_1M_2=SP$, }  \non \\
 A_3^{i}&=& \int\cdots \cases{ \left( r_\chi^{M_1}\Phi_{M_2}(x)\Phi_{m_1}(y)
 \,{2\bar y\over \bar xy(1-x\bar y)}+r_\chi^{M_2}\Phi_{M_1}(y)\Phi_{m_2}
 (x)\,{2x\over \bar xy(1-x\bar y)}\right); & for~$M_1M_2=PS$, \cr
 \left( -r_\chi^{M_1}\Phi_{M_2}(x)\Phi_{m_1}(y)
 \,{2\bar y\over \bar xy(1-x\bar y)}+r_\chi^{M_2}\Phi_{M_1}(y)\Phi_{m_2}
 (x)\,{2x\over \bar xy(1-x\bar y)}\right); & for~$M_1M_2=SP$,}
 \non \\
 A_3^{f} &=& \int\cdots \cases{ \left(r_\chi^{M_1}
 \Phi_{M_2}(x)\Phi_{m_1}(y)\,{2(1+\bar x)\over \bar x^2y}-r_\chi^{M_2}
 \Phi_{M_1}(y)\Phi_{m_2}(x)\,{2(1+y)\over \bar xy^2}\right); &
 for~$M_1M_2=PS$, \cr
 \left( -r_\chi^{M_1}
 \Phi_{M_2}(x)\Phi_{m_1}(y)\,{2(1+\bar x)\over \bar x^2y}-r_\chi^{M_2}
 \Phi_{M_1}(y)\Phi_{m_2}(x)\,{2(1+y)\over \bar xy^2}\right); &
 for~$M_1M_2=SP$,} \non \\
  A_1^f &=& A_2^f=0,
 \en
where $\int\cdots=\pi\alpha_s \int^1_0dxdy$, $\bar x=1-x$ and
$\bar y=1-y$. Note that we have adopted the same convention as in
\cite{BN} that $M_1$ contains an antiquark from the weak vertex
with longitudinal fraction $\bar y$, while $M_2$ contains a quark
from the weak vertex with momentum fraction $x$.

Using the asymptotic distribution amplitudes for pseudoscalar
mesons and keeping the LCDA of the scalar meson to the third
Gegenbaur polynomial in Eq. (\ref{eq:twist2wf}), the annihilation
contributions can be simplified to
 \be
 A_1^i(PS) &\approx& 2f_P f_S\pi\alpha_s\left\{9\mu_S \left[B_1(3X_A+4-\pi^2)
+B_3\left(10X_A+{23\over 18}-{10\over 3}\pi^2\right)\right]
-r_\chi^S r_\chi^P X_A^2\right\}, \non \\
 A_2^i(PS) &\approx& 2f_P f_S\pi\alpha_s\left\{-9\mu_S\left[B_1(X_A+29-3\pi^2)
+B_3\left(X_A+{2956\over 9}-{100\over 3}\pi^2\right)\right]
 +r_\chi^S r_\chi^P X_A^2\right\}, \non \\
 A_3^i(PS) &\approx& 6f_P f_S\pi\alpha_s \Bigg\{r_\chi^P \mu_S
 \left[3B_1\left(X_A^2-4X_A+4+{\pi^2\over 3}\right)+10B_3\left(X_A^2-
{19\over 3}X_A+{191\over 18}+{\pi^2\over 3}\right)\right]
 \non \\ &+& r_\chi^S\left(X_A^2-2X_A
 +{\pi^2\over 3}\right)\Bigg\}, \non \\
 A_3^f(PS) &\approx& 6f_P f_S\pi\alpha_s X_A\left\{r_\chi^P \mu_S
 \left[B_1(6X_A-11)+B_3\left(20X_A-{187\over 3}\right)\right]-r_\chi^S(2X_A-1) \right\},
 \en
for $M_1M_2=PS$, and
 \be
 A_1^i(SP)=A_2^i(PS),\qquad \qquad  A_2^i(SP)=A_1^i(PS), \non \\
  A_3^i(SP)=-A_3^i(PS), \qquad \qquad  A_3^f(SP)=A_3^f(PS),
 \en
for $M_1M_2=SP$, where the endpoint divergence $X_A$ is defined in
Eq. (\ref{eq:XA}). As noticed in passing, for neutral scalars
$\sigma$, $f_0$ and $a_0^0$, one needs to express $f_S r_\chi^S$
by $\bar f_S\bar r_\chi^S$ and $f_S\mu_S$ by $\bar f_S$.
Numerically, the dominant annihilation contribution arises from
the factorizable penguin-induced annihilation characterized by
$A_3^f$. Physically, this is because the penguin-induced
annihilation contribution is not subject to helicity suppression.

Although the parameters $a_i(i\neq 6,8)$ and $a_{6,8}r_\chi$ are
formally renormalization scale and $\gamma_5$ scheme independent,
in practice there exists some residual scale dependence in
$a_i(\mu)$ to finite order. To be specific, we shall evaluate the
vertex corrections to the decay amplitude at the scale
$\mu=m_b/2$. In contrast, as stressed in \cite{BBNS}, the hard
spectator and annihilation contributions should be evaluated at
the hard-collinear scale $\mu_h=\sqrt{\mu\Lambda_h}$ with
$\Lambda_h\approx 500 $ MeV. There is one more serious
complication about these contributions; that is, while QCD
factorization predictions are model independent in the
$m_b\to\infty$ limit, power corrections always involve troublesome
endpoint divergences. For example, the annihilation amplitude has
endpoint divergences even at twist-2 level and the hard spectator
scattering diagram at twist-3 order is power suppressed and posses
soft and collinear divergences arising from the soft spectator
quark. Since the treatment of endpoint divergences is model
dependent, subleading power corrections generally can be studied
only in a phenomenological way. We shall follow \cite{BBNS} to
parameterize the endpoint divergence $X_A\equiv\int^1_0 dx/\bar x$
in the annihilation diagram as
 \be \label{eq:XA}
 X_A=\ln\left({m_B\over \Lambda_h}\right)(1+\rho_A e^{i\phi_A}),
 \en
with the unknown real parameters $\rho_A$ and $\phi_A$. Likewise,
the endpoint divergence $X_H$ in the hard spectator contributions
can be parameterized in a similar manner.

Besides the penguin and annihilation contributions formally of
order $1/m_b$, there may exist other power corrections which
unfortunately cannot be studied in a systematical way as they are
nonperturbative in nature. The so-called ``charming penguin"
contribution is one of the long-distance effects that have been
widely discussed. The importance of this nonpertrubative effect
has also been conjectured to be justified in the context of
soft-collinear effective theory \cite{Bauer}. More recently, it
has been shown that such an effect can be incorporated in
final-state interactions \cite{CCS}. However, in order to see the
relevance of the charming penguin effect to $B$ decays into scalar
resonances, we need to await more data with better accuracy.

\begin{table}[t]
\caption{Branching ratios (in units of $10^{-6}$) of $B$ decays to
final states containing scalar mesons. The theoretical errors
correspond to the uncertainties due to (i) the Gegenbauer moments
$B_{1,3}$, the scalar meson decay constants, (ii) the
heavy-to-light form factors and the strange quark mass, and (iii)
the power corrections due to weak annihilation and hard spectator
interactions, respectively. The predicted branching ratios of
$B\to f_0(980)K,f_0(980)\pi$ are for the $f_0-\sigma$ mixing angle
$\theta=155^\circ$. For light scalar mesons $f_0(980)$, $a_0(980)$
and $\kappa$ we have assumed the 2-quark content for them. The
scalar mesons $a_0(1450)$ and $K_0^*(1450)$ are treated as the
first excited states of $a_0(980)$ and $\kappa$, respectively,
corresponding to scenario 1 explained in Appendices B and C.
Experimental results are taken from Table \ref{tab:exptBR}.}
\label{tab:theoryBR1}
\begin{ruledtabular}
\begin{tabular}{l r c |l  r c}
Mode & Theory & Expt & Mode & Theory & Expt  \\
\hline
 $B^-\to f_0(980)K^-$ & $15.6^{+0.3+4.7+5.4}_{-0.3-3.3-2.4}$ & $17.1^{+3.3}_{-3.5}$ &
 $\ov B^0\to f_0(980)\ov K^0$ & $13.3^{+0.2+4.1+4.5}_{-0.2-2.9-2.1}$  & $11.2\pm2.4$ \non \\
 $B^-\to f_0(980)\pi^-$ & $0.9^{+0.0+0.3+0.2}_{-0.0-0.2-0.0}$ & $<5.7$ &
 $\ov B^0\to f_0(980)\pi^0$ & $0.03^{+0.01+0.03+0.08}_{-0.01-0.00-0.01}$  &  \non \\
 $B^-\to a_0^0(980)K^-$ & $2.2^{+0.7+0.7+7.6}_{-0.5-0.5-1.7}$ & $<3.0$ & $\ov B^0\to a_0^+(980)K^-$ &
 $4.3^{+1.3+1.4+14.8}_{-1.1-1.0-~3.4}$ & $<1.9$  \non \\
 $B^-\to a_0^-(980)\ov K^0$ & $4.9^{+1.4+1.8+16.1}_{-1.1-1.2-~4.0}$ & $<4.6$ & $\ov B^0\to a_0^0(980)\ov K^0$
 & $2.4^{+0.7+0.9+7.9}_{-0.6-0.6-2.0}$ & $<9.2$ \non \\
 $B^-\to a_0^0(980)\pi^-$ & $3.4^{+0.2+1.0+0.4}_{-0.2-0.8-0.4}$ & $<6.9$ & $\ov B^0\to a_0^+(980)\pi^-$
 & $7.6^{+0.7+2.0+2.2}_{-0.6-1.8-1.6}$ & $<3.3$ \footnotemark[1]  \non \\
 $B^-\to a_0^-(980)\pi^0$ & $0.2^{+0.1+0.0+0.2}_{-0.1-0.0-0.1}$ &  & $\ov B^0\to a_0^-(980)\pi^+$
 & $0.6^{+0.2+0.1+0.7}_{-0.1-0.1-0.3}$ &  \non \\
 & & &  $\ov B^0\to a_0^0(980)\pi^0$ &
 $0.2^{+0.1+0.0+0.1}_{-0.1-0.0-0.0}$ & \non \\
 $B^-\to a_0^0(1450)K^-$ & $5.6^{+2.2+3.5+8.6}_{-1.7-1.9-5.2}$ &  & $\ov B^0\to a_0^+(1450)K^-$ &
 $11.1^{+4.4+6.9+17.1}_{-3.4-3.8-10.2}$ &  \non \\
 $B^-\to a_0^-(1450)\ov K^0$ & $14.1^{+5.0+8.2+18.9}_{-3.9-4.6-14.0}$ &  & $\ov B^0\to a_0^0(1450)\ov K^0$
 & $6.6^{+2.3+3.9+9.0}_{-1.9-2.2-6.6}$ &  \non \\
 $B^-\to a_0^0(1450)\pi^-$ & $4.1^{+0.5+1.1+1.3}_{-0.4-1.0-1.1}$ & & $\ov B^0\to a_0^+(1450)\pi^-$
 & $12.9^{+2.3+2.4+10.0}_{-2.0-2.2-~7.5}$ &  \non \\
 $B^-\to a_0^-(1450)\pi^0$ & $0.6^{+0.2+0.1+0.5}_{-0.2-0.1-0.3}$ &  & $\ov B^0\to a_0^-(1450)\pi^+$
 & $0.1^{+0.1+0.1+0.9}_{-0.1-0.0-0.0}$ &  \non \\
 & & &  $\ov B^0\to a_0^0(1450)\pi^0$ &
 $0.3^{+0.2+0.1+0.2}_{-0.1-0.1-0.1}$ & \non \\
 $B^-\to \ov K^{*0}_0(1430)\pi^-$ & $1.0^{+0.8+2.0+19.5}_{-0.5-0.7-~0.9}$ & $38.2^{+4.6}_{-4.5}$ &
 $\ov B^0\to K^{*-}_0(1430)\pi^+$  & $1.1^{+0.8+2.1+17.7}_{-0.5-0.9-~1.0}$ & $47.2^{+5.6}_{-6.9}$ \non \\
 $B^-\to K^{*-}_0(1430)\pi^0$ & $0.3^{+0.3+0.8+8.9}_{-0.2-0.2-0.3}$ &  &
 $\ov B^0\to \ov K^{*0}_0(1430)\pi^0$  & $0.6^{+0.4+1.0+8.8}_{-0.3-0.5-0.5}$ & $12.7\pm5.4$
\end{tabular}
\end{ruledtabular}
\footnotetext[1]{The cited upper limit $3.3\times 10^{-6}$ is for
$\ov B^0\to a_0^\pm(980)\pi^\mp$.}
\end{table}

\begin{table}[t]
\caption{Same as Table \ref{tab:theoryBR1} except that the mesons
$a_0(1450)$ and $K_0^*(1450)$ are treated as the lowest lying
scalar states, corresponding to scenario 2 explained in Appendices
B and C.} \label{tab:theoryBR2}
\begin{ruledtabular}
\begin{tabular}{l r c |l  r c}
Mode & Theory & Expt & Mode & Theory & Expt  \\
\hline
 $B^-\to a_0^0(1450)K^-$
 & $0.2^{+0.2+0.1+17.6}_{-0.0-0.1-~0.0}$ &  & $\ov B^0\to a_0^+(1450)K^-$
 & $0.3^{+0.5+0.3+36.4}_{-0.0-0.1-~0.0}$ &  \non \\
 $B^-\to a_0^-(1450)\ov K^0$
 & $0.1^{+0.6+0.3+35.9}_{-0.0-0.0-~0.1}$ &  & $\ov B^0\to a_0^0(1450)\ov K^0$
 & $0.1^{+0.3+0.2+17.7}_{-0.0-0.0-~0.0}$ \non \\
 $B^-\to a_0^0(1450)\pi^-$
 & $2.5^{+0.3+0.9+1.2}_{-0.3-0.7-0.8}$ &  & $\ov B^0\to a_0^+(1450)\pi^-$
 & $3.1^{+1.0+1.3+6.8}_{-0.9-1.0-1.8}$ \non \\
 $B^-\to a_0^-(1450)\pi^0$
 & $1.1^{+0.4+0.1+1.1}_{-0.3-0.1-0.6}$ & & $\ov B^0\to a_0^-(1450)\pi^+$
 & $0.5^{+0.2+0.2+2.6}_{-0.2-0.2-0.3}$ \non \\
 & & &  $\ov B^0\to a_0^0(1450)\pi^0$ &
 $0.7^{+0.3+0.1+0.5}_{-0.2-0.0-0.3}$ & \non \\
 $B^-\to \ov K^{*0}_0(1430)\pi^-$
 & $11.0^{+10.3+7.5+49.9}_{-~6.0-3.5-10.1}$ &  $38.2^{+4.6}_{-4.5}$ & $\ov B^0\to K^{*-}_0(1430)\pi^+$
 & $11.3^{+9.4+3.7+45.8}_{-5.8-3.7-~9.9}$ & $47.2^{+5.6}_{-6.9}$ \non \\
 $B^-\to K^{*-}_0(1430)\pi^0$
 & $5.3^{+4.7+1.6+22.3}_{-2.8-1.7-~4.7}$ & &  $\ov B^0\to \ov K^{*0}_0(1430)\pi^0$
 & $6.4^{+5.4+2.2+26.1}_{-3.3-2.1-~5.7}$ & $12.7\pm5.4$
\end{tabular}
\end{ruledtabular}
\end{table}

\subsection{Results and discussions}

While it is widely believed that $f_0(980)$ and $a_0(980)$ are
predominately four-quark states, in practice it is difficult to
make quantitative predictions on hadronic $B\to SP$ decays based
on the four-quark picture for light scalar mesons as it involves
not only the unknown form factors and decay constants  that are
beyond the conventional quark model but also additional
nonfactorizable contributions that are difficult to estimate (an
example will be shown shortly below). Hence, we shall assume the
two-quark scenario for $f_0(980)$ and $a_0(980)$.

For form factors we shall use those derived in the covariant
light-front quark model \cite{CCH}. For CKM matrix elements we use
the updated Wolfenstein parameters $A=0.825$, $\lambda=0.2262$,
$\bar\rho=0.207$ and $\bar\eta=0.340$ \cite{CKMfitter}. For the
running current quark masses we employ
 \be
 && m_b(m_b)=4.2\,{\rm GeV}, \qquad~~~ m_b(2.1\,{\rm GeV})=4.95\,{\rm
 GeV}, \qquad m_b(1\,{\rm GeV})=6.89\,{\rm
 GeV}, \non \\
 && m_c(m_b)=1.3\,{\rm GeV}, \qquad~~~ m_c(2.1\,{\rm GeV})=1.51\,{\rm
 GeV}, \non \\
 && m_s(2.1\,{\rm GeV})=90\,{\rm MeV}, \quad m_s(1\,{\rm GeV})=119\,{\rm
 MeV}, \non\\
 && m_d(1\,{\rm GeV})=6.3\,{\rm  MeV}, \quad~ m_u(1\,{\rm GeV})=3.5\,{\rm
 MeV}.
 \en
The strong coupling constants are given by
 \be
 \alpha_s(2.1\,{\rm GeV})=0.303,\qquad\quad \alpha_s(1\,{\rm
 GeV})=0.517\,,
 \en
corresponding to the world average $\alpha_s(m_Z)=0.1213$
\cite{PDG}.

The calculated results for branching ratios and \CP asymmetries
are exhibited in Tables
\ref{tab:theoryBR1}-\ref{tab:theoryCP2}.\footnote{$B$ decays into
light scalar mesons are not listed in Tables \ref{tab:theoryBR2}
and \ref{tab:theoryCP2} as we do not have a handle for light
scalars made of four quarks as explained in the text.}
In these tables we have included theoretical errors arising from
the uncertainties in the Gegenbauer moments $B_{1,3}$ (cf.
Appendix C), the scalar meson decay constant $f_S$ or $\bar f_S$
(see Appendix B), the form factors $F^{BP,BS}$, the quark masses
and the power corrections from weak annihilation and hard
spectator interactions characterized by the parameters $X_A$ and
$X_H$, respectively. For form factors we assign their
uncertainties to be $\delta F^{BP,BS}(0)=\pm 0.03$, for example,
$F_0^{BK}(0)=0.35\pm0.03$ and $F_0^{BK^*_0}(0)=0.26\pm0.03$. The
strange quark mass is taken to be $m_s(2\,{\rm GeV})=90\pm20$~MeV.
For the quantities $X_A$ and $X_H$ we adopt the form (\ref{eq:XA})
with $\rho_{A,H}\leq 0.5$ and arbitrary strong phases
$\phi_{A,H}$. Note that the central values (or ``default" results)
correspond to $\rho_{A,H}=0$ and $\phi_{A,H}=0$.

To obtain the errors shown in Tables
\ref{tab:theoryBR1}-\ref{tab:theoryCP2}, we first scan randomly
the points in the allowed ranges of the above six parameters in
three separated groups: the first two, the second two and the last
two, and then add errors in each group in quadrature. Therefore,
the first theoretical error shown in the Tables is due to the
variation of $B_{1,3}$ and $f_S$, the second error comes from the
uncertainties of the form factors and the strange quark mass,
while the third error from the power corrections due to weak
annihilation and hard spectator interactions.

Just like the $B$ decays into $PP$ or $VP$ final states in the QCD
factorization approach \cite{BBNS,BN}, the theoretical errors are
dominated by the $1/m_b$ power corrections due to weak
annihilation. However, it is clear from Tables
\ref{tab:theoryBR1}-\ref{tab:theoryBR2} that the theoretical
uncertainties in decay rates due to weak annihilation in some
$B\to SP$ decays, e.g. $B\to a_0(980)K,~a_0(1450)K$ and
$K^*_0(1430)\pi$ can be much larger than the ``default" central
values, while in $B\to PP$ or $VP$ decays, the errors due to
$X_{A,H}$ are comparable to or smaller than the central values
(see e.g. Table 2 of \cite{BN}). This can be understood as
follows. Consider the penguin-induced annihilation diagram for
$B\to PP$. Its amplitude is helicity suppressed as the helicity of
one of the final-state mesons cannot match with that of its
quarks. However, this helicity suppression can be alleviated in
the scalar meson production because of the non-vanishing orbital
angular momentum $L_z$ with the scalar state. Consequently, weak
annihilation contributions to $B\to SP$ can be much larger than
the $B\to PP$ case.

Finally, it is worth mentioning that we shall implicitly use the
narrow width approximation in the calculation of the $B$ decays
into resonances; that is, we will neglect the finite width effect
even for very broad resonances such as $\sigma$ and $\kappa$
states. Under the narrow width approximation, the resonant decay
rate respects a simple factorization relation (see e.g.
\cite{Chengf1370})
 \be
 \Gamma(B\to SP\to P_1P_2P)=\Gamma(B\to SP)\B(S\to P_1P_2).
 \en
It has been shown in \cite{Chengf1370} that in practice, this
factorization relation works reasonably well even for charmed
meson decays as long as the two-body decay $D\to SP$ is
kinematically allowed and the resonance is narrow. The off
resonance peak effect of the intermediate resonant state will
become important only when $D\to SP$ is kinematically barely or
even not allowed. The factorization relation presumably works much
better in $B$ decays due to its large energy release.

\subsubsection{$B\to f_0(980)K$ and $a_0(980)K$ decays}

\begin{figure}[t]
\vspace{-1cm}
  \centerline{\psfig{figure=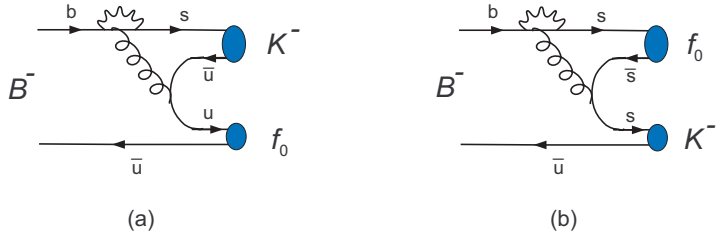,width=11cm}}
\vspace{0cm}
    \caption[]{\small Penguin contributions to $B^-\to f_0(980)K^-$.}
     \label{fig:BfK2q}
\end{figure}

The decay mode $B\to f_0(980)K$ has been studied in \cite{Chen}
within the framework of the pQCD approach based on the $k_T$
factorization theorem. It is found that the branching ratio is of
order $5\times 10^{-6}$ (see Fig. 2 of the second reference in
\cite{Chen}), which is smaller than the measured value by a factor
of 3.

The penguin-dominated $B\to f_0K$ decay receives two different
types of penguin contributions as depicted in Fig.
\ref{fig:BfK2q}. In the expression of $B\to f_0K$ decay amplitudes
given in Eq. (\ref{eq:SDAmp}), the superscript $u$ of the form
factor $F_0^{Bf_0^u}$ reminds us that it is the $u$ quark
component of $f_0$ involved in the form factor transition [Fig.
\ref{fig:BfK2q}(a)]. In contrast, the superscript $s$ of the decay
constant $\bar f_{f_0}^s$ indicates that it is the strange quark
content of $f_0$ responsible for the penguin contribution of Fig.
\ref{fig:BfK2q}(b). Note that $a_4$ and $a_6$ penguin terms
contribute constructively to $\pi^0K^-$ but destructively to $f_0
K^-$. Therefore, the contribution to $B\to f_0K$ from Fig. 1(a)
will be severely suppressed. Likewise, the contribution from Fig.
1(b) is suppressed by $\bar r_\chi^{f_0}\sim m_{f_0}/m_b$. Hence,
it is naively expected that the $f_0K$ rate is smaller than the
$\pi^0K$ one. However, as shown in Appendix B, the scale dependent
decay constant $\bar f_{f_0}^s$ is much larger than $f_\pi$ owing
to its scale dependence and the large radiative corrections to the
quark loops in the OPE series. As a consequence, the branching
ratio of $B\to f_0K$ turns out to be comparable to and even larger
than $B\to \pi^0K$.

Based on the QCD factorization approach, we obtain $\B(B^-\to
f_0K^-)=(9.0-13.5)\times 10^{-6}$ for $25^\circ< \theta<40^\circ$
and $(12.0-17.2)\times 10^{-6}$ for $140^\circ<\theta<165^\circ$
(Fig. 2), where only the central values are quoted.\footnote{The
calculated branching ratios in the present work are slightly
larger than that in \cite{CYf0K} because of the larger scalar
decay constant $\bar f_{f_0}^s$ and different estimates of the
leading-twist LCDA for $f_0(980)$. It was originally argued in
\cite{CYf0K} that while the extrinsic gluon contribution to $B\to
f_0K$ is negligible, the intrinsic gluon within the $B$ meson may
play an eminent role for the enhancement of $f_0(980)K$.}
Hence, the short-distance contributions suffice to explain the
observed large rates of $f_0K^-$ and $f_0\ov K^0$.

\begin{figure}[t]
\vspace{0cm} \centerline{\psfig{figure=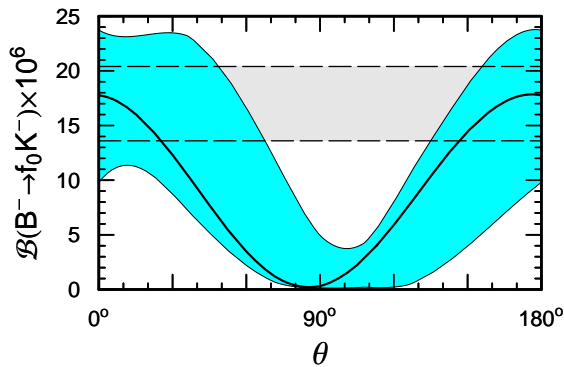,width=8cm}}
    \caption[]{\small The branching ratio of $B^-\to f_0(980)K^-$ versus the
    mixing angle $\theta$ of strange and nonstrange components of $f_0(980)$,
    where the middle bold solid curve inside the allowed region
    corresponds to the central value. For simplicity, theoretical errors due to
    weak annihilation and hard spectator interactions are not
    taken into account.
    The horizontal band within the dashed lines
    shows the experimentally allowed region with one sigma error.}
\end{figure}

Thus far we have discussed $f_0K$ modes with the two-quark
assignment for the $f_0(980)$. It is natural to ask what will
happen if $f_0$ is a four-quark bound state. Naively, one may
wonder if the energetic $f_0(980)$ produced in $B$ decays is
dominated by the four-quark configuration as it requires to pick
up two energetic quark-antiquark pairs to form a fast-moving light
four-quark scalar meson. The Fock states of $f_0(980)$ consist of
$q\bar q$, $q^2\bar q^2$, $q\bar q g,\cdots,$ etc.  It is thus
expected that the distribution amplitude of $f_0$ would be smaller
in the four-quark model than in the two-quark picture. Naively,
the observed $B\to f_0(980)K$ rates seem to imply that the
two-quark component of $f_0(980)$ play an essential role for this
weak decay.

\begin{figure}[t]
\vspace{0cm}
  \centerline{\psfig{figure=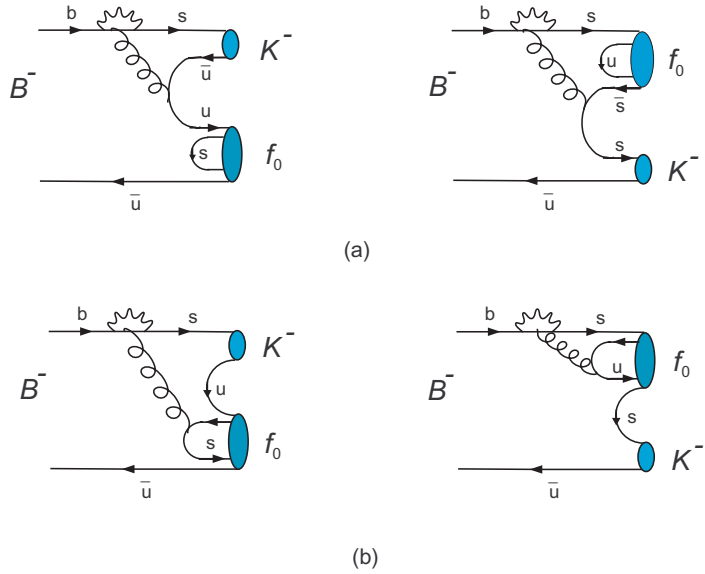,width=10cm}}
\vspace{0cm}
    \caption[]{\small Penguin contributions to $B^-\to
    f_0(980)K^-$in the 4-quark picture for $f_0(980)$.}
    \label{fig:BfK4q}
\end{figure}

Nevertheless, as pointed out in \cite{Brito}, the number of the
quark diagrams for the penguin contributions to $B\to f_0(980)K$
(Fig. \ref{fig:BfK4q}) in the four-quark scheme for $f_0(980)$ is
two times as many as that in the usual 2-quark picture (Fig.
\ref{fig:BfK2q}). That is, besides the factorizable diagrams in
Fig. \ref{fig:BfK4q}(a), there exist two more nonfactorizable
contributions depicted in Fig. \ref{fig:BfK4q}(b). Therefore, {\it
a priori} there is no reason that the $B\to f_0(980)K$ rate will
be suppressed if $f_0$ is a four-quark state. However, in
practice, it is difficult to give quantitative predictions based
on this scenario as the nonfactorizable diagrams are usually not
amenable. Moreover, even for the factorizable contributions, the
calculation of the $f_0(980)$ decay constant and its form factors
is beyond the conventional quark model, though an attempt has been
made in \cite{Brito}. In order to make quantitative calculations
for $B\to f_0(980)K$, we have assumed the conventional 2-quark
description of the light scalar mesons. However, as explained
before, the fact that its rate can be accommodated in the 2-quark
picture for $f_0(980)$ does not mean that the measurement of $B\to
f_0K$ can be used to distinguish between the 2-quark and 4-quark
assignment for $f_0(980)$.

We next turn to $B\to a_0(980)K$ decays. A main difference between
$a_0^0K$ and $f_0K$ modes is that the latter receives the dominant
contribution from the $s$ quark component of the $f_0$ [see Fig.
\ref{fig:BfK2q}(b)], while such a contribution vanishes in the
former mode even when $a_0^0$ is assigned with the $s\bar s(u\bar
u-d\bar d)/\sqrt{2}$ quark content. Because of the destructive
interference between the $a_4$ and $a_6$ terms, the penguin
contributions related to the $u$ quark component of the $a_0$ and
$f_0$ are largely suppressed. Consequently, the weak annihilation
contribution becomes as important as the penguin one. For example,
the branching ratio of $a_0^0 K^-$ is of order $3.3\times 10^{-7}$
in the absence of weak annihilation, while it becomes $2.4\times
10^{-6}$ when weak annihilation is turned on. From Table
\ref{tab:theoryBR1} we see that $\Gamma(B\to a_0^0K)\ll
\Gamma(B\to f_0K)$ and the $a_0^\pm K$ rate is enhanced by a
factor of 2 for charged $a_0$. The predicted central value of
$\B(\ov B^0\to a_0^+K^-)$ is larger than the current upper limit
by a factor of 2.  However, one cannot conclude definitely at this
stage that the 2-quark picture for $a_0(980)$ is ruled out since
it is still consistent with experiment when theoretical
uncertainties are taken into account. Nevertheless, as we shall
see below, when the unknown parameter $\rho_A$ for weak
annihilation is fixed to be of order 0.7 in order to accommodate
the $K_0^*(1430)\pi$ data, this in turn implies too large
$a_0(980)K$ rates compared to experiment. There will be more about
this when we discuss $B\to K^*_0(1430)\pi$ decays. Note that the
prediction of $\B(B^-\to a_0^-(980)\ov K^0)=15\times 10^{-6}$ made
in \cite{Minkowski} in the absence of the gluonic component is
ruled out by experiment.

\subsubsection{$B\to a_0(980)\pi,f_0(980)\pi$ decays}

The tree dominated decays $B\to a_0(980)\pi,f_0(980)\pi$ are
governed by the $B\to a_0$ and $B\to f_0^u$ transition form
factors, respectively. The $f_0\pi$ rate is rather small because
of the small $u\bar u$ component in the $f_0(980)$ and the
destructive interference between $a_4$ and $a_6$ penguin terms.
Since the $B\to a_0(980)$ form factor is predicted to be similar
to that for $B\to\pi$ one according to the covariant light-front
model (see Sec. III.C), it is interesting to compare $B\to a_0\pi$
decays with $B\to\pi\pi$. First, $\ov B^0\to a_0^-\pi^+$ is highly
suppressed. This means that the $B^0-\ov B^0$ interference plays
no role in the $a_0^\pm\pi^\mp$ channels. Thus the decays $\ov
B^0\to a_0^\pm\pi^\mp$ are expected to be self-tagging; that is,
the charge of the pion identifies the $B$ flavor. Second, we see
from Table \ref{tab:theoryBR1} that the branching ratio $\B(\ov
B^0\to a_0^+\pi^-)\sim 7.6\times 10^{-6}$ is slightly larger than
$\B(\ov B^0\to\pi^+\pi^-)=(4.5\pm0.4)\times 10^{-6}$ \cite{HFAG}
and that $\B(B^-\to \pi^-\pi^0)>\B(B^-\to a_0^0\pi^-)> \B(B^-\to
a_0^-\pi^0)$.

\begin{figure}[t]
\vspace{0cm}
  \centerline{\psfig{figure=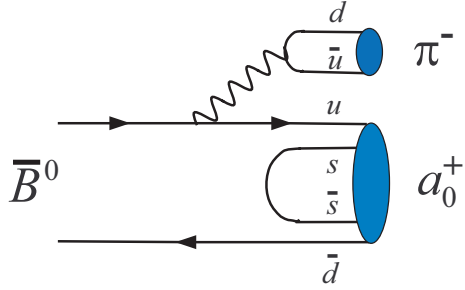,width=7cm}}
\vspace{0cm}
    \caption[]{\small Tree contribution to $\ov B^0\to
    a_0^+(980)\pi^-$in the 4-quark picture for $a_0(980)$.}
     \label{fig:a0pi}
\end{figure}

Just as the $a_0^+K^-$ mode, the predicted branching ratio $\B(\ov
B^0\to
a_0^\pm(980)\pi^\mp)=(8.2^{+0.9+2.1+2.9}_{-0.7-1.9-1.9})\times
10^{-6}$ exceeds the current experimental limit of $3.3\times
10^{-6}$ by more than a factor of 2 (cf. Table
\ref{tab:theoryBR1}). If the measured rate of $a_0^+\pi^-$ is at
the level of $(1\sim 2)\times 10^{-6}$ or even smaller, this will
imply a substantially smaller $B\to a_0$ form factor than the
$B\to\pi$ one. Hence, the four-quark explanation of the $a_0$ (see
Fig. \ref{fig:a0pi}) is preferred to account for the $B\to a_0$
form factor suppression. We shall see later that since $a_0(1450)$
can be described by the $q\bar q$ quark model, the study of
$a_0^+(1450)\pi^-$ relative to $a_0^+(980)\pi^-$ can provide a
more strong test on the quark content of $a_0(980)$. It has been
claimed in \cite{Suzuki} that the positive identification of
$B^0/\ov B^0\to a_0^\pm(980)\pi^\mp$ is an evidence against the
four-quark assignment of $a_0(980)$ or else for breakdown of
perturbative QCD. We disagree and we argue below that if the
branching ratio of $\ov B^0\to a_0^+(1450)\pi^-$ is measured at
the level of $3\times 10^{-6}$ and the $a_0^+(980)\pi^-$ rate is
found to be smaller, say, of order $(1\sim 2)\times 10^{-6}$, it
will be likely to imply a 2-quark nature for $a_0(1450)$ and a
four-quark assignment for $a_0(980)$.

In short, although it is unlikely that the penguin-dominated decay
$B\to f_0K$ can be used to distinguish between the 2-quark and
4-quark assignment for $f_0(980)$, the decays $B\to a_0\pi$ and
$a_0K$ may serve for the same purpose for $a_0(980)$. For example,
the former mode is tree dominated and its amplitude is
proportional to the form factor $F_0^{Ba_0}$ which is suppressed
in the four-quark model for $a_0(980)$. It has been claimed in
\cite{Delepine} that a best candidate to distinguish the nature of
the $a_0$ scalar is $\B(B^-\to a_0^-\pi^0)$ as the prediction for
a four-quark model is one order of magnitude smaller than for the
two quark assignment. We see from Table \ref{tab:theoryBR1} that
the branching ratio of this mode is only of order $2\times
10^{-7}$ even when $a_0(980)$ is treated as a 2-quark state.
Experimentally, it would be extremely difficult to test the
$a_0(980)$ nature from the study of $a_0^-(980)\pi^0$.

It is commonly assumed that only the valence quarks of the initial
and final state hadrons participate in the decays. Nevertheless, a
real hadron in QCD language should be described by a set of Fock
states for which each state has the same quantum number as the
hadron. For example,
\begin{eqnarray}\label{eq:fockexpansion}
|a^+(980)\rangle &=& \psi_{u\bar d}^{a_0} |u\bar d\rangle +
\psi_{u\bar dg}^{a_0} |u\bar d g\ra + \psi_{u\bar d s\bar s}^{a_0}
|u\bar d s \bar s\rangle+ \dots\,.
\end{eqnarray}
The possibility that $a_0(980)$ can be viewed as a bound state of
four quarks at low energies, while its 2-quark component manifests
at high energies is also allowed by current experiments.

Note that the production of $a_0(980)$ in hadronic $B$ decays has
not been seen so far and only some limits have been set. In
contrast, the $a_0(980)$ production in charm decays has been
measured in several places, e.g. $D^0\to K^0a_0^0(980)$ and
$K^-a_0^+(980)$ in the three-body decays $D^0\to K^+K^-\ov K^0$
\cite{BaBar3K}. It is conceivable that the scalar resonance
$a_0(980)$ in $B$ decays will be seen at $B$ factories soon.

\subsubsection{$B\to K_0^*(1430)\pi$ decays}

For weak decays involving scalar mesons above 1 GeV such as
$K_0^*(1430)$, $a_0(1450)$ and $f_0(1500)$ we consider two
different scenarios to evaluate their decay constants and LCDAs
based on the QCD sum rule method (see Appendices B and C): i)
$K_0^*(1430), a_0(1450), f_0(1500)$ are treated as the first
excited states of $\kappa, a_0(980)$ and $f_0(980)$, respectively,
and (ii) they are the lowest lying resonances and the
corresponding first excited states lie in between $(2.0\sim
2.3)$~GeV. Scenario 2 corresponds to the case that light scalar
mesons are four-quark bound states, while all scalar mesons are
made of two quarks in scenario 1. The resultant decay constants
and LCDAs for the scalar mesons above 1 GeV in these two different
scenarios are summarized in Appendices B and C. The $B\to
K^*_0(1430)$ form factors in scenarios 1 and 2 can be found in
Table \ref{tab:FF}. It should be stressed that the decay constants
of $K_0^*(1430), a_0(1450), f_0(1500)$ have the signs flipped from
scenario 1 to scenario 2 as explained in footnote 9 in Appendix B.

As mentioned in the Introduction, there exists a two-fold
experimental ambiguity in extracting the branching ratio of
$B^-\to \ov K_0^*(1430)^0\pi^-$: Belle found two different
solutions for its branching ratios from the fit to $B^+\to
K^+\pi^+\pi^-$ events \cite{BelleKpipi}. The larger solution is
consistent with BaBar \cite{BaBarKpipi} while the other one is
smaller by a factor of 5 [see Eq. (\ref{eq:K*0pi+})]. It appears
that the larger of the two solutions, namely, $\B(B^-\to \ov
K_0^*(1430)^0\pi^-)\sim 45\times 10^{-6}$ , is preferable as it is
consistent with the BaBar measurement and supported by a
phenomenological estimate in \cite{Chernyak}. However, since
$B^-\to \ov K^0\pi^-$ has a branching ratio of order $24\times
10^{-6}$ \cite{HFAG}, one may wonder why the $\ov K_0^{*0}\pi^-$
production is much more favorable than $\ov K^0\pi^-$, while the
$\ov K^{*0}_0\pi^0$ mode is comparable to $\ov K^0\pi^0$ (see
Table \ref{tab:exptBR}).

To proceed we consider the pure penguin decays $B^-\to \ov
K^{*0}_0\pi^-$ and $B^-\to \ov K^0\pi^-$ for the purpose of
illustration. The dominant penguin amplitudes read [see also Eq.
(\ref{eq:SDAmp})]
 \be
 A(B^-\to \ov K^{*0}_0\pi^-) &\propto&
 (a_4^p-r_\chi^{K^*_0}a_6^p)_{\pi
 K_0^*}\,f_{K^*_0}F_0^{B\pi}(m^2_{K^*_0})(m_B^2-m_\pi^2),
 \non \\
  A(B^-\to \ov K^0\pi^-) &\propto&
 (a_4^p+r_\chi^Ka_6^p)_{\pi
 K}\,f_{K}F_0^{B\pi}(m^2_{K})(m_B^2-m_\pi^2),
 \en
where we have neglected annihilation contributions for the time
being. Although the decay constant of $K^*_0(1430)$, which is
$37\pm4$ MeV in scenario 2 [cf. Eq.
(\ref{app:decayconstantsK02})],  is much smaller than that of the
kaon, it is compensated by the large ratio $r_\chi^{K_0^*}=8.9$ at
$\mu=2.1$ GeV compared to $r_\chi^K=1.1$. Since the penguin
coefficient $a_6$ is the same for both $K_0^*\pi$ and $K\pi$
modes, it is thus expected that $\Gamma(\ov
K^{*0}_0\pi^-)/\Gamma(\ov K^0\pi^-)\approx 3.2$ in the absence of
the $a_4$ contribution. When $a_4$ is turned on, we notice that
its contribution is destructive to $\ov K^{*0}_0\pi^-$ and
constructive to $\ov K^0\pi^-$. In order to see the effect of
$a_4$ explicitly we give the numerical results for the relevant
$a_i^p(\pi K_0^*)$ at the scale $\mu=2.1$ GeV
 \be \label{eq:ai(piK0*)}
  && a_1 = 1.417 - i0.181, \qquad\qquad\quad~~ a_2=0.673-i0.111,
  \non \\
  &&  a_4^u = -0.199 - i0.009, \qquad\qquad\quad
  a_4^c = -0.162 - i0.059, \non \\
  && a_6^u = -0.0558 -i 0.0163, \qquad\qquad
  a_6^c = -0.0602 -i 0.0039, \non \\
  && a_8^u = (79.4 - i4.8)\times 10^{-5}, \qquad\quad~
  a_8^c = (78.5 - i2.4)\times 10^{-5},  \non \\
  && a_{10}^u = (70 -i64)\times 10^{-4}, \qquad\qquad
  a_{10}^c = (70 - i62)\times 10^{-4},
  \en
and for $a_i^p(\pi K)$
 \be \label{eq:ai(piK)}
  && a_1 = 0.993 + i0.0288, \qquad\qquad\quad~ a_2=0.144-i0.111,
  \non \\
  &&  a_4^u = -0.0267 - i0.0183, \qquad\qquad
  a_4^c = -0.0343 - i0.0064, \non \\
  && a_6^u = -0.0568 -i 0.0163, \qquad\qquad
  a_6^c = -0.0612 -i 0.0039, \non \\
  && a_8^u = (74.4 - i4.5)\times 10^{-5}, \qquad\quad~
  a_8^c = (73.6 - i2.3)\times 10^{-5},  \non \\
  && a_{10}^u = (-208 + i90)\times 10^{-5}, \quad\quad~~
  a_{10}^c = (-209 + i90)\times 10^{-5},
  \en
where scenario 2 has been used to evaluate both $a_i^p(\pi K_0^*)$
and $a_i^p(\pi K)$. Comparing Eq. (\ref{eq:ai(piK0*)}) and with
Eq. (\ref{eq:ai(piK)}), it is evident that vertex and spectator
interaction corrections to $a_1,a_2,a_4$ and $a_{10}$ for $\pi
K^*_0$ are quite large compared to the corresponding $a_i(\pi K)$
due mainly to the different nature of the $K^*_0$ LCDA. Note that
the $a_6$ and $a_8$ penguin terms remain intact as they do not
receive vertex and hard spectator interaction contributions. Since
the magnitude of $a_4^p(\pi K_0^*)$ is increased significantly, it
is clear that the $\ov K_0^{*0}\pi^-$ rate eventually becomes
slightly smaller than $\ov K^0\pi^-$ due to the large destructive
contribution from $a_4^p(\pi K^*_0)$. Hence, we conclude that
$\B(B^-\to\ov K_0^{*0}\pi^-)\sim 1\times 10^{-5}\sim{1\over 2}
\B(B^-\to\ov K^0\pi^-)$ in the absence of weak annihilation
contributions.

>From Tables \ref{tab:theoryBR1} and \ref{tab:theoryBR2} it is
clear that when weak annihilation is turned on, the $K^*_0\pi$
rates are highly suppressed in scenario 1 due to the large
destructive contributions from the defaulted weak annihilation. In
order to accommodate the data, one has to take into account the
power corrections due to the non-vanishing $\rho_A$ and $\rho_H$
from weak annihilation and hard spectator interactions,
resepctively. Since power corrections are dominated by weak
annihilation, a fit to the data yields $\rho_A\sim 0.4$ for
scenario 2 and $\rho_A\sim 0.7$ for scenario 1, where we have
taken $\phi_A\approx 0$.

We see from Eq. (\ref{eq:SDAmp}) that the amplitudes of $B^-\to
\ov K^{*0}_0\pi^-$ and $\ov B^0\to K^{*-}_0\pi^+$ are identical
when the small contributions from the electroweak penguin and
$\lambda_u a_1$, $\lambda_ub_2$ terms are neglected. This amounts
to assuming  the dominance of the $\Delta I=0$ penguin
contributions. Hence, these two modes should have the same rates
under the isospin approximation \cite{Gronau}. Likewise,
$\Gamma(\ov B^0\to \ov K^{*0}_0\pi^0)/\Gamma(\ov B^0\to
K^{*-}_0\pi^+)=1/2$ is expected to hold in the isospin limit.
Indeed, it is found in QCD factorization calculations
that\footnote{From Table \ref{tab:theoryBR1} and Eq.
(\ref{eq:ratios}), it appears that the mode $K^{*-}_0\pi^0$ does
not respect the approximated isospin relation $\Gamma(B^-\to
K^{*-}_0\pi^0)/\Gamma(\ov B^0\to K^{*-}_0\pi^+)=1/2$. This is
mainly ascribed to the large cancellation between penguin and
annihilation terms in the amplitude of $B^-\to K^{*-}_0\pi^0$ [see
Eq. (\ref{eq:SDAmp})] and the remaining term proportional to
$(a_2\delta_u^p+3(a_9-a_7)/2)$ breaks isospin symmetry.}
 \be \label{eq:ratios}
 R_1\equiv {\B(\ov B^0\to \ov K_0^{*0}(1430)\pi^0)\over \B(\ov B^0\to
 K_0^{*-}(1430)\pi^+)} &=& \cases{0.51^{+0.01+0.08+0.04}_{-0.02-0.02-0.06}
 & scenario~1; \cr 0.47^{+0.01+0.02+0.04}_{-0.02-0.01-0.10} & scenario~2;} \non \\
 R_2\equiv {\B(B^-\to K_0^{*-}(1430)\pi^0)\over \B(B^-\to \ov
 K_0^{*0}(1430)\pi^-)} &=& \cases{0.30^{+0.03+0.43+0.79}_{-0.02-0.02-0.02}
 & scenario~1; \cr 0.58^{+0.05+0.01+0.17}_{-0.03-0.01-0.05} & scenario~2;} \\
 R_3\equiv {\tau(B^0)\B(B^-\to \ov K_0^{*0}(1430)\pi^-)\over \tau(B^-)\B(\ov B^0\to
 K_0^{*-}(1430)\pi^+)} &=& \cases{0.81^{+0.06+0.62+0.93}_{-0.07-0.04-0.57}
 & scenario~1; \cr 0.90^{+0.05+0.03+0.18}_{-0.06-0.03-0.27} & scenario~2.} \non
 \en
Consequently, the ambiguity in regard to $B^-\to \ov
K_0^{*0}\pi^-$ found by Belle can be resolved by the measurement
of $\ov B^0\to K^{*-}_0\pi^+$. As noted in passing, both BaBar and
Belle measurements of $B^-\to \ov K^{*0}_0\pi^-$ and $B^0\to
K^{*-}_0\pi^+$ [see Eq. (\ref{eq:K*0pi+}) and Table
\ref{tab:exptBR}] do respect the isospin relation. It is also
important to measure the ratio of $\B(\ov B^0\to \ov
K_0^{*0}(1430)\pi^0)/\B(\ov B^0\to
 K_0^{*-}(1430)\pi^+)$ to see if it is close to one half.
At any rate, both BaBar and Belle should measure all
$K^*_0(1430)\pi$ modes with a careful Dalitz plot analysis of
nonresonant contributions to three-body decays to avoid any
possible ambiguities.

We now turn to the implications of sizable weak annihilation
characterized by the parameter $\rho_A$ which is of order 0.7 in
scenario 1 and ${\cal O}(0.4)$ in scenario 2. We find that all the
calculated $a_0(980)K$ rates are too large compared to experiment.
For example, $\B(\ov B^0\to a_0^+(980)K^-)\approx 31.4\times
10^{-6}$ for $\rho_A=0.7$ and $\approx 14.6\times 10^{-6}$ for
$\rho_A=0.4$. Both are ruled out by the current limit of
$1.9\times 10^{-6}$. This clearly indicates that $a_0(980)$ cannot
be a purely two-quark state and that scenario 2 in which the light
scalar meson is assigned to be a four-quark state is preferable.

\subsubsection{$B\to a_0(1450)K,a_0(1450)\pi$ decays}

For $\ov B\to a_0(1450)\pi$ and $a_0(1450)K$ decays, the
calculated results should be reliable as the $a_0(1450)$ can be
described by the $q\bar q$ quark model. Just as $a_0(980)K$ modes,
weak annihilation gives a dominant contribution to $a_0(1450)K$
rates. It is found that their rates are much larger in scenario 1
than in scenario 2 due to the relative sign difference between the
Gegenbauer moments $B_1$ and $B_3$ for $a_0(1450)$ and the sign of
the $a_0(1450)$ decay constant flipped in these two scenarios (see
Tables \ref{tab:momentscenario1} and \ref{tab:momentscenario2}).
The interference pattern between the penguin and annihilation
amplitudes is generally opposite in scenarios 1 and 2. For
example, the interference in $\ov B^0\to a_0^+(1450)K^-$ is
constructive in scenario 1 but becomes destructive in scenario 2.
By the same token, the $a_0^+(1450)\pi^-$ and $a_0^0(1450)\pi^0$
rates are also quite different in scenarios 1 and 2.

As discussed in the previous subsection on $K_0^*(1430)\pi$,
predictions under scenario 2 are more preferable. Hence, if the
branching ratio of $\ov B^0\to a_0^\pm(1450)\pi^\mp$ is measured
at the level of $4\times 10^{-6}$ and the $a_0^+(980)\pi^-$ rate
is found to be smaller, say, of order $(1\sim 2)\times 10^{-6}$ or
even smaller than this, it will be likely to imply a 2-quark
nature for $a_0(1450)$ and a four-quark assignment for $a_0(980)$.
Note that the naive estimate of $20\times 10^{-6}$ made by
\cite{Chernyak} for this mode appears to be too large due to the
usage of a large $B\to a_0(1450)$ form factor,
$F_0^{Ba_0(1450)}(0)=0.46$. Experimentally, $a_0(1450)$ will be
more difficult to identify than $a_0(980)$ because of its broad
width, $265\pm 13$ MeV \cite{PDG}.

\subsubsection{$\ov B^0\to\kappa^+ K^-$ as spectroscope for
$\kappa$ four quark state}

\begin{figure}[t]
\vspace{-1.5cm}
  \centerline{\psfig{figure=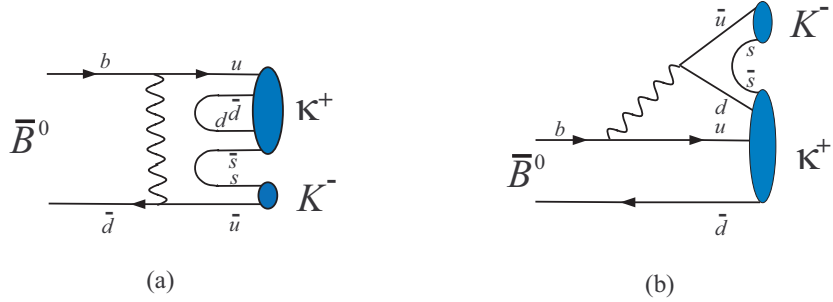,width=13cm}}
\vspace{0cm}
    \caption[]{\small Annihilation and tree contributions to $\ov B^0\to
    \kappa^+K^-$ in the 4-quark picture for $\kappa$.}
     \label{fig:kappaK}
\end{figure}

As for $\kappa$ (or $K_0^*(800)$), there is a nice and unique
place where one can discriminate between the 4-quark and 2-quark
pictures for the $\kappa$ meson, namely, the $\ov B^0\to \kappa^+
K^-$ decay. Recall that $\ov B^0\to K^+K^-$ is strongly suppressed
as it can only proceed through the $W$-exchange diagram. The
experimental upper bound on its branching ratio is $0.6\times
10^{-6}$ \cite{HFAG,PDG} while it is predicted to be of order
$1\times 10^{-8}$ theoretically (see e.g. \cite{BN}). Naively $\ov
B^0\to \kappa^+ K^-$ is also rather suppressed if $\kappa$ is made
of two quarks. However, if $\kappa$ has primarily a four-quark
content, this decay can receive a tree contribution as depicted in
Fig. \ref{fig:kappaK}(b). Hence, if $\ov B^0\to \kappa^+ K^-$ is
observed at the level of $\gsim 10^{-7}$, it may imply a
four-quark content for the $\kappa$. Presumably, this can be
checked from the Dalitz plot analysis of the three-body decay $\ov
B^0\to K^+K^-\pi^0$ or $\ov B^0\to K^0K^-\pi^+$. As noticed
before, scenario 2 is more favored for explaining the $B\to
K^*_0(1430)\pi$ data. This already implies that $\kappa$ is
preferred to be a four-quark state.

Unlike the other light scalar mesons, the experimental evidence
for $\kappa$  is still controversial. The $\kappa$ state has been
reported by E791 in the analysis of $D^+\to K^-\pi^+\pi^+$ with
the mass $797\pm19\pm43$ MeV and width $410\pm43\pm87$ MeV
\cite{E791}. However, CLEO did not see evidence for the $\kappa$
in $D^0\to K^-\pi^+\pi^0$ \cite{CLEOkappa}. The $\kappa$ state was
also reported by the reanalyses of LASS data on $\pi K$ scattering
phase shifts using the $T$-matrix method \cite{Bugg} and the
unitarization method combined with chiral symmetry \cite{Zheng}.
Most recently, BES has reported the evidence for the $\kappa$ in
$J/\psi\to \ov K^{*0}K^+\pi^-$ process with the mass
$878\pm23^{+64}_{-55}$ MeV and width $499\pm52^{+55}_{-87}$ MeV
\cite{BESII}.

It is interesting to notice that the decays $\ov B^0\to D_s^+K^-$
and $\ov B^0\to D_{s0}^*(2317)^+ K^-$, the analogues of $\ov
B^0\to K^+K^-$ and $\ov B^0\to \kappa^+ K^-$, have been measured
recently. The measured branching ratios are $\B(\ov B^0\to
D_s^+K^-)=(3.8\pm1.3)\times 10^{-5}$ \cite{PDG} and $\B(\ov B^0\to
D_{s0}^{*+}K^-)\B(D_{s0}^{*+}\to D_s^+\pi^0)=(5.3^{+1.5}_{-1.3}\pm
1.6)\times 10^{-6}$ \cite{BelleDs0K}. Since $D_{s0}^*(2317)^+$ is
dominated by the hadronic decay into $D_s^+\pi^0$, it is clear
that $\Gamma(\ov B^0\to D_{s0}^{*+}K^-)\gsim \Gamma(\ov B^0\to
D_s^+K^-)$. These two decays can only proceed via a short-distance
$W$-exchange process or through the long-distance final-state
rescattering processes $\ov B^0\to D^+\pi^-\to D_s^+ K^-$ and $\ov
B^0\to D^{*+}_0\pi^-\to D^{*+}_{s0} K^-$. (In fact, the
rescattering process has the same topology as $W$-exchange.) Since
$\B(\ov B^0\to D^+\pi^-)\approx 2.8\times 10^{-3}\gg \B(\ov B^0\to
D_s^+K^-)$, it is thus expected that the decay $\ov B^0\to D_s^+
K^-$  is dominated by the long-distance rescattering process. As
$\B(\ov B^0\to D_0^{*+}\pi^-)<1.8\times 10^{-4}$ \cite{PDG}, we
will naively conclude that $\Gamma(\ov B^0\to
D_{s0}^{*+}K^-)/\Gamma(\ov B^0\to D_s^+K^-) < 0.06$, in
contradiction to the experimental observation. Nevertheless, if
$D_{s0}^*(2317)^+$ is a bound state of $c\bar sd\bar d$~\cite{CH},
then a tree diagram similar to Fig. \ref{fig:kappaK}(b) will
contribute and this may allow us to explain why $\Gamma(\ov B^0\to
D_{s0}^{*+}K^-)\gsim \Gamma(\ov B^0\to D_s^+K^-)$.

\subsubsection{$B\to\sigma\pi$ decays} The tree dominated
$B\to\sigma\pi$ decays are expected to have similar rates as
$B\to\pi^0\pi$ ones if the $\sigma$ meson is assumed to be a bound
state of 2 quarks. Assuming that $\sigma$ has similar decay
constant and LCDA as $f_0(980)$, it is found that
$\B(B^-\to\sigma\pi^-)\approx 4.5\times 10^{-6}$ and $\B(\ov
B^0\to\sigma\pi^0)\approx 1.7\times 10^{-7}$. The former is to be
compared with the upper limit $4.1\times 10^{-6}$ \cite{BaBar3pi}.

\begin{table}[t]
\caption{Same as Table \ref{tab:theoryBR1} except for \CP
asymmetries (in \%).}
 \label{tab:theoryCP1}
\begin{ruledtabular}
\begin{tabular}{l r c |l  r c}
Mode & Theory & Expt & Mode & Theory & Expt  \\
\hline
 $B^-\to f_0(980)K^-$ & $0.4^{+0.0+0.0+0.6}_{-0.0-0.0-0.6}$ & $-2.0^{+6.8}_{-6.5}$ &
 $\ov B^0\to f_0(980)\ov K^0$ & $0.7^{+0.0+0.0+0.1}_{-0.0-0.0-0.1}$  & $-6\pm21$ \non \\
 $B^-\to f_0(980)\pi^-$ & $-2.0^{+0.2+0.1+43.0}_{-0.2-3.4-41.9}$ & $-50\pm54$  &
 $\ov B^0\to f_0(980)\pi^0$ & $38.6^{+11.8+~2.9+~44.1}_{-11.0-18.3-114.4}$  &  \non \\
 $B^-\to a_0^0(980)K^-$ & $3.8^{+1.6+1.5+52.8}_{-1.1-1.5-63.6}$ &  & $\ov B^0\to a_0^+(980)K^-$ &
 $3.4^{+1.4+1.4+51.3}_{-1.0-1.4-61.9}$ & \non \\
 $B^-\to a_0^-(980)\ov K^0$ & $0.9^{+0.1+0.1+1.2}_{-0.1-0.2-0.9}$ &  & $\ov B^0\to a_0^0(980)\ov K^0$ &
 $0.7^{+0.1+0.1+1.0}_{-0.1-0.1-0.5}$ & \non \\
 $B^-\to a_0^0(980)\pi^-$ & $-0.6^{+0.1+0.1+3.6}_{-0.1-0.2-3.8}$ &  & $\ov B^0\to a_0^+(980)\pi^-$
 & $-0.3^{+0.2+0.5+23.6}_{-0.2-0.3-22.8}$ &   \non \\
 $B^-\to a_0^-(980)\pi^0$ & $-65.9^{+6.1+7.8+24.6}_{-8.3-5.2-22.0}$ &  & $\ov B^0\to a_0^-(980)\pi^+$
 & $76.5^{+~5.9+4.0+21.0}_{-10.5-5.9-36.1}$ &  \non \\
 & & & $\ov B^0\to a_0^0(980)\pi^0$ &
 $34.3^{+12.3+~9.1+28.6}_{-~8.3-11.6-30.5}$ & \non \\
 $B^-\to a_0^0(1450)K^-$ & $0.9^{+0.5+0.9+21.4}_{-0.3-0.6-18.7}$ &  & $\ov B^0\to a_0^+(1450)K^-$
 & $0.9^{+0.5+0.9+19.7}_{-0.3-0.6-18.9}$ & \non \\
 $B^-\to a_0^-(1450)\ov K^0$ & $0.3^{+0.1+0.1+0.2}_{-0.1-0.1-6.6}$ &  & $\ov B^0\to a_0^0(1450)\ov K^0$ &
 $0.3^{+0.1+0.1+0.2}_{-0.0-0.1-1.7}$ & \non \\
 $B^-\to a_0^0(1450)\pi^-$ & $-2.9^{+0.2+0.4+5.4}_{-0.2-0.3-5.7}$ &  & $\ov B^0\to a_0^+(1450)\pi^-$
 & $0.4^{+0.1+0.5+36.2}_{-0.1-0.3-35.4}$ &  \non \\
 $B^-\to a_0^-(1450)\pi^0$ & $19.8^{+1.8+3.2+44.9}_{-6.4-3.6-46.6}$ &  & $\ov B^0\to a_0^-(1450)\pi^+$
 & $59.2^{+13.8+12.6+~33.0}_{-36.7-66.6-152.7}$ & \non \\
 & & & $\ov B^0\to a_0^0(1450)\pi^0$ &
 $-32.8^{+11.7+7.8+66.2}_{-16.7-6.6-52.4}$ & \non \\
 $B^-\to \ov K^{*0}_0(1430)\pi^-$ & $-4.4^{+2.8+~4.1+63.8}_{-4.9-25.7-30.5}$ & $-5^{+5}_{-8}$ &
 $\ov B^0\to K^{*-}_0(1430)\pi^+$  & $15.1^{+5.0+18.5+16.1}_{-3.6-~5.9-21.2}$ & $-7\pm14$ \non \\
 $B^-\to K^{*-}_0(1430)\pi^0$ & $-42.1^{+12.6+78.6+128.7}_{-10.1-~2.1-~12.3}$ &  &
 $\ov B^0\to \ov K^{*0}_0(1430)\pi^0$  & $3.4^{+0.4+0.3+10.8}_{-0.3-7.4-~9.1}$ & $-34\pm19$ \non \\
\end{tabular}
\end{ruledtabular}
\end{table}

\begin{table}[t]
\caption{Same as Table \ref{tab:theoryBR2} except for \CP
asymmetries (in \%).}
 \label{tab:theoryCP2}
\begin{ruledtabular}
\begin{tabular}{l r c |l  r c}
Mode & Theory & Expt & Mode & Theory & Expt  \\
\hline
 $B^-\to a_0^0(1450)K^-$ & $54.7^{+~2.3+~3.6+~10.4}_{-33.1-16.2-101.0}$ &
 & $\ov B^0\to a_0^+(1450)K^-$
 & $53.3^{+~4.8+~5.3+10.0}_{-35.0-18.3-96.4}$ & \non \\
 $B^-\to a_0^-(1450)\ov K^0$
 & $19.9^{+~4.1+~8.8+8.4}_{-17.0-13.5-4.5}$ &  & $\ov B^0\to a_0^0(1450)\ov K^0$
 & $4.9^{+0.6+0.5+1.9}_{-6.5-6.2-6.2}$ & \non \\
 $B^-\to a_0^0(1450)\pi^-$
 & $-0.9^{+0.2+0.1+6.3}_{-0.2-0.3-6.6}$ & & $\ov B^0\to a_0^+(1450)\pi^-$
 & $-0.7^{+0.0+0.3+46.9}_{-0.0-0.2-45.2}$ \non \\
 $B^-\to a_0^-(1450)\pi^0$
 & $-41.4^{+6.7+1.0+123.7}_{-6.2-1.7-~64.8}$ & & $\ov B^0\to a_0^-(1450)\pi^+$
 & $37.6^{+10.0+1.3+48.6}_{-11.9-2.1+68.0}$ \non \\
 & & &  $\ov B^0\to a_0^0(1450)\pi^0$ &
 $20.8^{+7.5+3.3+43.3}_{-4.9-4.0-45.6}$ & \non \\
 $B^-\to \ov K^{*0}_0(1430)\pi^-$ & $1.1^{+1.3+0.2+~8.0}_{-0.7-0.1-17.8}$  & $-5^{+5}_{-8}$ &
 $\ov B^0\to K^{*-}_0(1430)\pi^+$  & $-3.8^{+1.9+0.3+~8.3}_{-3.6-0.3-13.2}$ & $-7\pm14$ \non \\
 $B^-\to K^{*-}_0(1430)\pi^0$ & $4.2^{+3.6+1.2+5.0}_{-2.9-1.2-3.7}$ &  &
 $\ov B^0\to \ov K^{*0}_0(1430)\pi^0$  & $0.6^{+0.5+0.1+0.7}_{-0.3-0.1-0.7}$ & $-34\pm19$
\end{tabular}
\end{ruledtabular}
\end{table}

\begin{table}[t]
\caption{Mixing-induced \CP parameter $\Delta S\equiv
\sin2\beta_{\rm eff}-\sin2\beta_{\rm CKM}$ in scenarios 1 and 2 as
explained in Appendices B and C. The sources of theoretical errors
are same as in previous tables except the last one is from the
uncertainty in the unitarity angle $\gamma$.} \label{tab:theoryS}
\begin{ruledtabular}
\begin{tabular}{l r r }
Mode & Theory (Scenario 1)  & Theory (Scenario 2) \\
\hline
 $\ov B {}^0\to f_0^0(980)K_S$
 & $0.023^{+0.000+0.000+0.001+0.001}_{-0.000-0.000-0.001-0.001}$
 &
 \non \\
 $\ov B {}^0\to a_0^0(980)K_S$
 & $0.022^{+0.000+0.000+0.005+0.001}_{-0.000-0.000-0.006-0.001}$
 &
 \non \\
 $\ov B {}^0\to a_0^0(1450)K_S$
 & $0.023^{+0.000+0.000+0.027+0.001}_{-0.000-0.000-0.001-0.001}$
 & $0.021^{+0.014+0.008+0.031+0.001}_{-0.000-0.000-0.009-0.001}$
 \non \\
 $\ov B {}^0\to \ov K^{*0}_0(1430)\pi^0$
 & $0.004^{+0.005+0.010+0.030+0.000}_{-0.008-0.040-0.036-0.000}$
 & $0.021^{+0.001+0.000+0.004+0.001}_{-0.002-0.013-0.008-0.001}$
\end{tabular}
\end{ruledtabular}
\end{table}

\subsubsection{Direct \CP asymmetries}

We see from Tables \ref{tab:theoryCP1} and \ref{tab:theoryCP2}
that \CP partial rate asymmetries in those charmless $B\to SP$
decays with branching ratios $\gsim 10^{-6}$ are in general at
most a few percents. This is ascribed to the fact that the strong
phases calculable in QCD factorization are generally small and
that the observation of direct \CP violation requires at least two
different contributing amplitudes with distinct strong and weak
phases. Hence, if the observed direct \CP asymmetry is of order
${\cal O}(0.1)$ or larger, then strong phases induced from power
corrections could be important. As pointed out in \cite{CCS},
final-state rescattering processes can have important effects on
the decay rates and their direct \CP violation, especially for
color-suppressed and penguin-dominated modes. However, this is
beyond the scope of the present work.

\subsubsection{Mixing-induced \CP asymmetries}

It is of great interest to measure the mixing-induced indirect \CP
asymmetries $S_f$ for penguin-dominated modes and compare them to
the one inferred from the charmonium mode ($J/\psi K_S$) in $B$
decays. It is expected in the Standard Model that $\sin2\beta_{\rm
eff}$ defined via $S_f\equiv -\eta_f\sin2\beta_{\rm eff}$ with
$\eta_f$ being the \CP eigenvalue of the final state $f$ should be
equal to $S_{J/\psi K_S}$ with a small deviation at most ${\cal
O}(0.1)$ \cite{LS}. See \cite{sin2beta} for recent studies of
$\sin2\beta_{\rm eff}$ in some of $B\to PP$ and $PV$ modes using
the QCD factorization approach with or without the presence of
final state interactions. In Table~\ref{tab:theoryS} we show the
predictions on the mixing-induced \CP parameter $\Delta S\equiv
\sin2\beta_{\rm eff}-\sin2\beta_{\rm CKM}$ for the \CP eigenstates
$f_0(980) K_S$, $a^0_0(980) K_S$, $a^0_0(1450) K_S$ and $\ov
K^{*0}_0(1430) \pi^0$, where  only the \CP component of $\ov
K^{*0}_0(1430)$ namely, $K_S\pi^0$, is considered in the last
mode. In addition to the theoretical errors considered before, the
uncertainty of $7^\circ$ in the unitarity angle $\gamma$ is
included. Note that main errors arise from the uncertainties in
annihilation contributions and $\gamma$. Our results indicate that
$\Delta S_f$ in these penguin dominated modes are positive and
very small.

\section{Conclusions}

In this work we have studied the hadronic $B$ decays into a scalar
meson and a pseudoscalar meson  within the framework of QCD
factorization. Vertex corrections, hard spectator interactions and
weak annihilation contributions to the hadronic $B\to SP$ decays
are studied using the QCD factorization approach. Our main results
are as follows:

\begin{itemize}


\item Based on the QCD sum rule method, we have derived the
leading-twist light-cone distribution amplitudes (LCDAs) of scalar
mesons and their decay constants. It is found that the scalar
decay constant is much larger than the previous estimates owing to
its scale dependence and the large radiative corrections to the
quark loops in the OPE series. Unlike the pseudoscalar or vector
mesons, the scalar LCDAs are governed by the odd Gegenbauer
polynomials.

\item While it is widely believed that light scalar mesons such as
$f_0(980)$, $a_0(980)$, $\kappa$ are predominately four-quark
states, in practice it is difficult to make quantitative
predictions on $B\to SP$ based on the four-quark picture for $S$
as it involves not only the form factors and decay constants that
are beyond the conventional quark model but also additional
nonfactorizable contributions that are difficult to estimate.
Hence, in practice we shall assume the two-quark scenario for
light scalar mesons in calculations.

\item The short-distance approach suffices to explain the observed
large rates of $f_0K^-$ and $f_0\ov K^0$ that receive major
penguin contributions from the penguin process $b\to ss\bar s$ and
are governed by the large $f_0$ scalar decay constant. When
$f_0(980)$ is assigned as a four-quark bound state, there exist
two times more diagrams contributing to $B\to f_0(980)K$.
Therefore, although the $f_0(980)K$ rates can be accommodated in
the 2-quark picture for $f_0(980)$, it does not mean that the
measurement of $B\to f_0K$ can be used to distinguish between the
2-quark and 4-quark assignment for $f_0(980)$.

\item When $a_0(980)$ is treated as a $q\bar q$ bound state, it is
found that the predicted $\ov B^0\to a_0^+(980)\pi^-$ and
$a_0^+(980)K^-$ rates exceed substantially the current
experimental limits. Hence, a four-quark assignment for $a_0(980)$
is favored. The $a_0(980)K$ and $a_0(1450)K$ receive dominant
contributions from weak annihilation.

\item Belle found two different solutions for the branching ratios
of $B^+\to K_0^*(1430)^0\pi^+$ from the fit to $B^+\to
K^+\pi^+\pi^-$ events. The larger solution is consistent with
BaBar while the other one is smaller by a factor of 5. Based on
the isospin argument, we have shown that the smaller of the two
solutions is ruled out by the measurements of $K_0^*(1430)^-\pi^+$
by BaBar and Belle.

\item For $B\to a_0(1450)\pi,~a_0(1450)K$ and $K^*_0(1430)\pi$
decays, we have explored two possible scenarios for the scalar
mesons above 1 GeV in the QCD sum rule method, depending on
whether the light scalars $\kappa,~a_0(980)$ and $f_0(980)$ are
treated as the lowest lying $q\bar q$ states or four-quark
particles. We pointed out that in both scenarios, one needs
sizable weak annihilation in order to accommodate the $K^*_0\pi$
data. This in turn implies that all the predicted $a_0(980)K$
rates in scenario 1 will be too large compared to the current
limits if $a_0(980)$ is a bound state of two quarks. This means
that the scenario in which the scalar mesons above 1 GeV are
lowest lying $q\bar q$ scalar state and the light scalar mesons
are four-quark states is preferable. The branching ratio of $\ov
B^0\to a_0^\pm(1450)\pi^\mp$ is predicted to be at the level of
$4\times 10^{-6}$.

\item The decay $\ov B^0\to \kappa^+ K^-$ can be used to
discriminate between the 4-quark and 2-quark nature for the
$\kappa$ meson. This mode is strongly suppressed if $\kappa$ is
made of two quarks as it can proceed through the $W$-exchange
process. However, if $\kappa$ is predominately a four-quark state,
it will receive a color-allowed tree contribution.  Hence, an
observation of this channel at the level of $\gsim 10^{-7}$ would
mostly imply a four-quark picture for the $\kappa$. Presumably,
this can be checked from the Dalitz plot analysis of three-body
decay $\ov B^0\to K^+K^-\pi^0$ or $\ov B^0\to K^0K^-\pi^+$.

\item Direct $CP$ asymmetries in those decay modes with branching
ratios $\gsim 10^{-6}$ are usually small of order a few percents.
However, final-state rescattering processes can have important
impact on the decay rates and their direct \CP violation.

\item Mixing-induced \CP asymmetries in the penguin dominated $SP$
modes such as $f_0(980) K_S$, $a^0_0(980) K_S$, $a^0_0(1450) K_S$
and $\ov K^{*0}_0(1430)[K_S\pi^0] \pi^0$ are studied. Their
deviations from $\sin2\beta_{\rm CKM}$ are found to be positive
($\Delta S>0$) and tiny.

\end{itemize}

\vskip 3.0cm \acknowledgments

This research was supported in part by the National Science
Council of R.O.C. under Grant Nos. NSC93-2112-M-001-043,
NSC93-2112-M-001-053 and NSC93-2112-M-033-004.

\vspace{3cm}

\appendix

\section{Decay amplitudes of $B\to SP$}

The $B\to SP$ ($PS$) decay amplitudes can be either evaluated
directly or obtained readily from $B\to VP$ ($PV$) amplitudes with
the replacements: $f_V\Phi_V(x)\to \Phi_S(x)$ and
$m_Vf_V^\bot\Phi_v(x)\to -m_S\Phi_S^s(x)$. (The factor of $i$ will
be taken care of by the factorizable amplitudes of $B\to SP$ shown
below.) To make the replacements more transparent, it is
convenient to employ the LCDA $\Phi_S(x)$ in the form
(\ref{eq:twist2wf}) and factor out the decay constants $f_S$ in
$\Phi_S(x)$ and $\bar f_S$ in $\Phi_S^s(x)$ [see Eq.
(\ref{eq:twist3wf})],  so that we have \footnote{We found in the
present work that it is most suitable to define the LCDAs of
scalar mesons including decay constants. In this appendix we try
to make connections between $B\to SP$ and $B\to VP$ amplitudes.
The latter have been worked out in detail in \cite{BN}. Since the
LCDAs in \cite{BN} are defined with the decay constants being
excluded, for our purposes it is more convenient to factor out the
decay constants in the scalar LCDAs so that it is ready to obtain
$B\to SP$ amplitudes from $B\to VP$ ones via the replacement
(\ref{eq:replacementI}).}
 \be \label{eq:replacementI}
 \quad \Phi_V(x)\to \Phi_S(x), \qquad \Phi_v(x)\to \Phi_S^s(x),
 \qquad f_V\to f_S, \qquad r^V_\chi\to -r_\chi^S,
 \en
where
 \be \label{eq:rchiS}
 r_\chi^V(\mu)={2m_V\over m_b(\mu)}\,{f_V^\bot(\mu)\over f_V}, \qquad\quad
 r_\chi^S(\mu)={2m_S\over m_b(\mu)}\,{\bar f_S(\mu)\over f_S}={2m_S^2\over m_b(\mu)(m_2(\mu)-m_1(\mu))},
 \en
and use of Eq. (\ref{eq:EOM}) has been made. For the neutral
scalars $\sigma$, $f_0$ and $a_0^0$, $r_\chi^S$ becomes divergent
while $f_S$ vanishes.  In this case one needs to express
$f_Sr_\chi^S$ by $\bar f_S\bar r_\chi^S$ with
 \be
 \bar r_\chi^S(\mu)={2m_S\over m_b(\mu)}.
 \en

With the above-mentioned replacements, the quantity $A_{M_1M_2}$
and the coefficients of the flavor operators $\alpha_i^p$ defined
in \cite{BN} read
 \be \label{eq:replacementII}
 A_{M_1M_2} &=& {G_F\over
 \sqrt{2}}\cases{(m_B^2-m_P^2)F_0^{BP}(m_S^2)f_S; & for~$M_1M_2=PS$,  \cr
 -(m_B^2-m_S^2)F_0^{BS}(m_P^2)f_P; & for~$M_1M_2=SP$, } \non \\
 \alpha_3^p(M_1M_2) &=& \cases{ a_3^p(M_1M_2)+a_5^p(M_1M_2); & for ~$M_1M_2=PS$, \cr
 a_3^p(M_1M_2)-a_5^p(M_1M_2); & for ~$M_1M_2=SP$,} \non \\
 \alpha_4^p(M_1M_2) &=& \cases{ a_4^p(M_1M_2)-r_\chi^S a_6^p(M_1M_2); & for ~$M_1M_2=PS$, \cr
 a_4^p(M_1M_2)-r_\chi^P a_6^p(M_1M_2); & for ~$M_1M_2=SP$,}  \\
 \alpha_{\rm 3,EW}^p(M_1M_2) &=& \cases{ a_9^p(M_1M_2)+a_7^p(M_1M_2); & for ~$M_1M_2=PS$, \cr
 a_9^p(M_1M_2)-a_7^p(M_1M_2); & for ~$M_1M_2=SP$,} \non \\
  \alpha_{\rm 4,EW}^p(M_1M_2) &=& \cases{ a_{10}^p(M_1M_2)-r_\chi^S a_8^p(M_1M_2); & for ~$M_1M_2=PS$, \cr
 a_{10}^p(M_1M_2)-r_\chi^P a_8^p(M_1M_2); & for ~$M_1M_2=SP$, }
 \non
 \en
where
 \be \label{eq:rchiP}
 r_\chi^P={2m_P^2\over m_b(\mu)(m_2+m_1)(\mu)}.
 \en
It should be stressed that the $a_i^pF_0^{BP}$ and $a_i^pF_0^{BS}$
terms in the decay amplitudes have an opposite sign.

Applying the replacement (\ref{eq:replacementI}) and Eq.
(\ref{eq:replacementII}) to the $B\to VP$ and $PV$ amplitudes
given in Appendix of \cite{BN}, we obtain the following the
factorizable amplitudes of the decays $B\to
f_0K,~a_0\pi,~\sigma\pi,~a_0K,~K_0^*\pi$
 \be \label{eq:SDAmp}
A(B^- \to f_0 K^- ) &=&
-\frac{G_F}{\sqrt{2}}\sum_{p=u,c}\lambda_p^{(s)}
 \Bigg\{ \left(a_1 \delta^p_u+a_4^p-r_\chi^Ka_6^p
 +a_{10}^p-r_\chi^Ka_8^p \right)_{f_0^u K} \non \\
 &\times& f_KF_0^{Bf_0^u}(m_K^2)(m_B^2-m_{f_0}^2)
 + \left(a_6^p-{1\over 2}a_8^p\right)_{Kf_0^s}\bar r_\chi^{f_0}\bar f^s_{f_0}\,F_0^{BK}
 (m^2_{f_0})(m_B^2-m_K^2)\non \\
 &-& f_B\bigg[\big(b_2\delta_u^p+b_3
 +b_{\rm 3,EW}\big)_{f_0^uK} +\big(b_2\delta_u^p+b_3
 +b_{\rm 3,EW}\big)_{Kf_0^s}\bigg] \Bigg\}, \non \\
A(\ov B^0 \to f_0\ov K^0 ) &=&
-\frac{G_F}{\sqrt{2}}\sum_{p=u,c}\lambda_p^{(s)}
 \Bigg\{ \left(a_4^p-r_\chi^Ka_6^p
 -{1\over 2}(a_{10}^p-r_\chi^Ka_8^p) \right)_{f_0^dK} \non \\
 &\times& f_KF_0^{Bf_0^u}(m_K^2)(m_B^2-m_{f_0}^2)
 + \left(a_6^p-{1\over 2}a_8^p\right)_{Kf_0^s}\bar r_\chi^{f_0}\bar
 f^s_{f_0}\,F_0^{BK}(m^2_{f_0})(m_B^2-m_K^2)\non \\
  &-& f_B\bigg[\big(b_3
 -{1\over 2}b_{\rm 3,EW}\big)_{f_0^dK} +\big(b_3
 -{1\over 2}b_{\rm 3,EW}\big)_{Kf_0^s}\bigg] \Bigg\}, \non \\
A(B^- \to a_0^0 K^- ) &=&
-\frac{G_F}{{2}}\sum_{p=u,c}\lambda_p^{(s)}
 \Bigg\{ \left(a_1 \delta^p_u+a_4^p-r_\chi^Ka_6^p
 +a_{10}^p-r_\chi^Ka_8^p \right)_{a_0 K} \non \\
 &\times& f_KF_0^{Ba_0}(m_K^2)(m_B^2-m_{a_0}^2)
 - f_B\big(b_2\delta_u^p+b_3
 +b_{\rm 3,EW}\big)_{a_0K}\Bigg\}, \non \\
A(B^- \to a_0^- \ov K^0 ) &=&
-\frac{G_F}{\sqrt{2}}\sum_{p=u,c}\lambda_p^{(s)}
 \Bigg\{ \left(a_4^p-r_\chi^Ka_6^p
 -{1\over 2}(a_{10}^p-r_\chi^Ka_8^p) \right)_{a_0 K} \non \\
&\times&  f_KF_0^{Ba_0}(m_K^2)(m_B^2-m_{a_0}^2)
 - f_B\big(b_2\delta_u^p+b_3
 +b_{\rm 3,EW}\big)_{a_0K} \Bigg\}, \non \\
A(\ov B^0 \to a_0^+ K^- ) &=&
-\frac{G_F}{\sqrt{2}}\sum_{p=u,c}\lambda_p^{(s)}
 \Bigg\{ \left(a_1\delta_u^p+a_4^p-r_\chi^Ka_6^p
 +a_{10}^p-r_\chi^Ka_8^p \right)_{a_0K} \non \\
&\times & f_KF_0^{Ba_0}(m_K^2)(m_B^2-m_{a_0}^2)
  - f_B\big(b_3
 -{1\over 2}b_{\rm 3,EW}\big)_{a_0K}\bigg] \Bigg\}, \non \\
A(\ov B^0 \to a_0^0\ov K^0 ) &=&
 \frac{G_F}{{2}}\sum_{p=u,c}\lambda_p^{(s)}
 \Bigg\{ \left(a_4^p-r_\chi^Ka_6^p
 -{1\over 2}(a_{10}^p-r_\chi^Ka_8^p) \right)_{a_0K} \non \\
 &\times& f_KF_0^{Ba_0}(m_K^2)(m_B^2-m_{a_0}^2)
  - f_B\big(b_3
 -{1\over 2}b_{\rm 3,EW}\big)_{a_0K} \Bigg\}, \non \\
A(B^- \to f_0 \pi^- ) &=&
-\frac{G_F}{\sqrt{2}}\sum_{p=u,c}\lambda_p^{(s)}
 \Bigg\{ \left(a_1 \delta^p_u+a_4^p-r_\chi^\pi a_6^p
 +a_{10}^p-r_\chi^\pi a_8^p \right)_{f_0^u \pi} \non \\
 &\times& f_\pi F_0^{Bf_0^u}(m_\pi^2)(m_B^2-m_{f_0}^2)
 + \left(a_6^p-{1\over 2}a_8^p\right)_{\pi f_0^u}\bar r_\chi^{f_0}\bar f^u_{f_0}\,F_0^{B\pi}
 (m^2_{f_0})(m_B^2-m_\pi^2)\non \\
 &-& f_B\bigg[\big(b_2\delta_u^p+b_3
 +b_{\rm 3,EW}\big)_{f_0^u \pi} +\big(b_2\delta_u^p+b_3
 +b_{\rm 3,EW}\big)_{\pi f_0^u}\bigg] \Bigg\}, \non \\
A(\ov B^0 \to f_0\pi^0 ) &=&
\frac{G_F}{2}\sum_{p=u,c}\lambda_p^{(s)}
 \Bigg\{ \left(-a_2 \delta^p_u+a_4^p-r_\chi^\pi a_6^p-{3\over 2}(a_9^p-a_7^p)
 -{1\over 2}(a_{10}^p-r_\chi^\pi a_8^p) \right)_{f_0^d \pi} \non \\
 &\times& f_\pi F_0^{B f_0^d}(m_\pi^2)(m_B^2-m_{f_0}^2)
 + \left(a_6^p-{1\over 2}a_8^p\right)_{\pi f_0^d}\bar r_\chi^{f_0}\bar
 f^d_{f_0}\,F_0^{B\pi}(m^2_{f_0})(m_B^2-m_\pi^2)\non \\
  &+& f_B\bigg[\big(b_1 \delta^p_u-b_3
 +{1\over 2}b_{\rm 3,EW}+{3\over 2}b_{\rm 4,EW}\big)_{f_0^d \pi} +\big(b_1 \delta^p_u-b_3
 +{1\over 2}b_{\rm 3,EW}+{3\over 2}b_{\rm 4,EW}\big)_{\pi f_0^d}\bigg] \Bigg\},\non \\
A(B^- \to \sigma_0 \pi^- ) &=&
-\frac{G_F}{\sqrt{2}}\sum_{p=u,c}\lambda_p^{(s)}
 \Bigg\{ \left(a_1 \delta^p_u+a_4^p-r_\chi^\pi a_6^p
 +a_{10}^p-r_\chi^\pi a_8^p \right)_{\sigma_0^u \pi} \non \\
 &\times& f_\pi F_0^{B\sigma_0^u}(m_\pi^2)(m_B^2-m_{\sigma_0}^2)
 + \left(a_6^p-{1\over 2}a_8^p\right)_{\pi \sigma_0^u}\bar r_\chi^{\sigma_0}\bar f^u_{\sigma_0}\,F_0^{B\pi}
 (m^2_{\sigma_0})(m_B^2-m_\pi^2)\non \\
 &-& f_B\bigg[\big(b_2\delta_u^p+b_3
 +b_{\rm 3,EW}\big)_{\sigma_0^u \pi} +\big(b_2\delta_u^p+b_3
 +b_{\rm 3,EW}\big)_{\pi \sigma_0^u}\bigg] \Bigg\}, \non \\
A(\ov B^0 \to \sigma_0\pi^0 ) &=&
\frac{G_F}{2}\sum_{p=u,c}\lambda_p^{(s)}
 \Bigg\{ \left(-a_2 \delta^p_u+a_4^p-r_\chi^\pi a_6^p-{3\over 2}(a_9^p-a_7^p)
 -{1\over 2}(a_{10}^p-r_\chi^\pi a_8^p) \right)_{\sigma_0^d \pi} \non \\
 &\times& f_\pi F_0^{B \sigma_0^d}(m_\pi^2)(m_B^2-m_{\sigma_0}^2)
 + \left(a_6^p-{1\over 2}a_8^p\right)_{\pi \sigma_0^d}\bar r_\chi^{\sigma_0}\bar
 f^d_{\sigma_0}\,F_0^{B\pi}(m^2_{f_0})(m_B^2-m_\pi^2)\non \\
  &+& f_B\bigg[\big(b_1 \delta^p_u-b_3
 +{1\over 2}b_{\rm 3,EW}+{3\over 2}b_{\rm 4,EW}\big)_{\sigma_0^d \pi} +\big(b_1 \delta^p_u-b_3
 +{1\over 2}b_{\rm 3,EW}+{3\over 2}b_{\rm 4,EW}\big)_{\pi \sigma_0^d}\bigg] \Bigg\},\non \\
  A(\ov B^0 \to a^+_0\pi^- ) &=&
-\frac{G_F}{\sqrt{2}}\sum_{p=u,c}\lambda_p^{(d)}
 \Bigg\{ \left( a_1\delta_u^p+ a_4^p-r_\chi^\pi a_6^p
 +a_{10}^p-r_\chi^\pi a_8^p \right)_{a_0\pi} \non \\
 &\times& f_\pi F_0^{Ba_0}(m_\pi^2)(m_B^2-m_{a_0}^2)
 - f_B\Big[\big(b_3+b_4
 -{1\over 2}b_{\rm 3,EW}-{1\over 2}b_{\rm 4,EW}\big)_{a_0\pi} \non
 \\  &+& \big(b_1\delta_u^p+b_4 +b_{\rm 4,EW}\big)_{\pi a_0} \Big]
 \Bigg\}, \non \\
A(\ov B^0 \to a^-_0\pi^+ ) &=&
 \frac{G_F}{\sqrt{2}}\sum_{p=u,c}\lambda_p^{(d)}
 \Bigg\{ \left( a_1\delta_u^p+ a_4^p-r_\chi^{a_0} a_6^p
 +a_{10}^p-r_\chi^{a_0} a_8^p \right)_{\pi a_0} \non \\
 &\times& f_{a_0} F_0^{B\pi}(m_\pi^2)(m_B^2-m_\pi^2)
 + f_B\Big[\big(b_3+b_4
 -{1\over 2}b_{\rm 3,EW}-{1\over 2}b_{\rm 4,EW}\big)_{\pi a_0} \non
 \\  &+& \big(b_1\delta_u^p+b_4 +b_{\rm 4,EW}\big)_{a_0\pi} \Big]
 \Bigg\}, \non \\
  A(B^- \to a^0_0\pi^- ) &=&
-\frac{G_F}{2}\sum_{p=u,c}\lambda_p^{(d)}
 \Bigg\{ \left( a_1\delta_u^p+ a_4^p-r_\chi^\pi a_6^p
 +a_{10}^p-r_\chi^\pi a_8^p \right)_{a_0\pi} \non \\
 &\times& f_\pi F_0^{Ba_0}(m_\pi^2)(m_B^2-m_{a_0}^2)
 -\left(a_6^p-{1\over 2}a_8^p\right)_{\pi a_0}\bar r_\chi^{a_0}\bar
 f_{a_0} (m_B^2-m_\pi^2)F_0^{B\pi}(m_{a_0}^2) \non \\
 &-& f_B\Big[\big(b_2\delta_\mu^p+b_3+
 b_{\rm 3,EW}\big)_{a_0\pi}-
 \big(b_2\delta_\mu^p+b_3 +b_{\rm 3,EW}\big)_{\pi a_0} \Big]
 \Bigg\}, \non \\
  A(B^- \to a^-_0\pi^0) &=&
 \frac{G_F}{2}\sum_{p=u,c}\lambda_p^{(d)}
 \Bigg\{ \left( a_1\delta_u^p+ a_4^p-r_\chi^{a_0} a_6^p
 +a_{10}^p-r_\chi^{a_0} a_8^p \right)_{\pi a_0}f_{a_0}
 F_0^{B\pi}(m_{a_0}^2)(m_B^2-m_\pi^2) \non \\
 &-& \left[a_2\delta_u^p-a_4^p+r_\chi^\pi a_6^p+{1\over 2}(a_{10}^p-r_\chi^\pi a_8^p)
 +{3\over 2}(a_9-a_7)\right]_{a_0\pi}f_\pi F_0^{Ba_0}(m_\pi^2)
 (m_B^2-m_{a_0}^2) \non \\
  &+& f_B\Big[\big(b_2\delta_\mu^p+b_3+
 b_{\rm 3,EW}\big)_{\pi a_0}-
 \big(b_2\delta_\mu^p+b_3 +b_{\rm 3,EW}\big)_{a_0\pi} \Big]
 \Bigg\}, \non \\
  A(\ov B^0 \to a^0_0\pi^0 ) &=&
-\frac{G_F}{2\sqrt{2}}\sum_{p=u,c}\lambda_p^{(d)}
 \Bigg\{ \left( a_2\delta_u^p- a_4^p+r_\chi^\pi a_6^p
 +{1\over 2}(a_{10}^p-r_\chi^\pi a_8^p)+{3\over 2}(a_9-a_7) \right)_{a_0\pi} \non \\
 &\times& f_\pi F_0^{Ba_0}(m_\pi^2)(m_B^2-m_{a_0}^2)
 -\left(a_6^p-{1\over 2}a_8^p\right)_{\pi a_0}\bar r_\chi^{a_0}\bar
 f_{a_0} (m_B^2-m_\pi^2)F_0^{B\pi}(m_{a_0}^2) \non \\
 &+& f_B\Big[\big(b_1\delta_\mu^p+b_3+2b_4-
 {1\over 2}(b_{\rm 3,EW}-b_{\rm 4,EW})\big)_{a_0\pi} \non \\
 &+& \big(b_1\delta_\mu^p+b_3+2b_4 - {1\over 2}(b_{\rm 3,EW}-b_{\rm 4,EW})
 \big)_{\pi a_0} \Big]
 \Bigg\}, \non \\
A(B^- \to \ov K^{*0}_0\pi^- ) &=&
\frac{G_F}{\sqrt{2}}\sum_{p=u,c}\lambda_p^{(s)}
 \Bigg\{ \left( a_4^p-r_\chi^{K^*_0}a_6^p
 -{1\over 2}(a_{10}^p-r_\chi^{K^*_0}a_8^p)\right)_{\pi K^*_0} \non \\
 &\times& f_{K_0^*}F_0^{B\pi}(m_{K_0^*}^2)(m_B^2-m_\pi^2)
 + f_B\big(b_2\delta_u^p+b_3
 +b_{\rm 3,EW}\big)_{\pi K^*_0} \Bigg\}, \non \\
A(B^- \to K^{*-}_0\pi^0 ) &=&
\frac{G_F}{2}\sum_{p=u,c}\lambda_p^{(s)}
 \Bigg\{ \left( a_1\delta_u^p+a_4^p-r_\chi^{K^*_0}a_6^p
 +a_{10}^p-r_\chi^{K^*_0}a_8^p \right)_{\pi K^*_0} \non \\
 &\times& f_{K_0^*}F_0^{B\pi}(m_{K_0^*}^2)(m_B^2-m_\pi^2)-\left[a_2\delta_u^p+{3\over
 2}(a_9-a_7)\right]_{K^*_0\pi}f_\pi F_0^{BK^*_0}(m_\pi^2)(m_B^2-m_{K_0^*}^2)
 \non \\
 &+& f_B\big(b_2\delta_u^p+b_3
 +b_{\rm 3,EW}\big)_{\pi K^*_0} \Bigg\}, \non \\
 A(\ov B^0 \to K^{*-}_0\pi^+ ) &=&
\frac{G_F}{\sqrt{2}}\sum_{p=u,c}\lambda_p^{(s)}
 \Bigg\{ \left( a_1\delta_u^p+ a_4^p-r_\chi^{K^*_0}a_6^p
 +a_{10}^p-r_\chi^{K^*_0}a_8^p \right)_{\pi K^*_0} \non \\
 &\times&
 f_{K_0^*}F_0^{B\pi}(m_{K_0^*}^2)(m_B^2-m_\pi^2)
 + f_B\big(b_3
 -{1\over 2}b_{\rm 3,EW}\big)_{\pi K^*_0} \Bigg\}, \non \\
 A(\ov B^0 \to \ov K^{*0}_0\pi^0 ) &=&
\frac{G_F}{2}\sum_{p=u,c}\lambda_p^{(s)}
 \Bigg\{ \left( -a_4^p+r_\chi^{K^*_0}a_6^p
 +{1\over 2}(a_{10}^p-r_\chi^{K^*_0}a_8^p) \right)_{\pi K^*_0} \non \\
 &\times&
 f_{K_0^*}F_0^{B\pi}(m_{K_0^*}^2)(m_B^2-m_\pi^2)-\left[a_2\delta_u^p+{3\over
 2}(a_9-a_7)\right]_{K^*_0\pi}f_\pi F_0^{BK^*_0}(m_\pi^2)(m_B^2-m_{K_0^*}^2)
 \non \\
 &+& f_B\big(-b_3
 +{1\over 2}b_{\rm 3,EW}\big)_{\pi K^*_0} \Bigg\}, 
 \en
where $\lambda_p^{(q)}\equiv V_{pb}V_{pq}^*$ with $q=d,s$ and
 \be \label{eq:r}
 && r^K_\chi(\mu)={2m_K^2\over m_b(\mu)(m_u(\mu)+m_s(\mu))}, \qquad
 r^{K^*_0}_\chi(\mu)={2m_{K_0^*}^2\over
 m_b(\mu)(m_s(\mu)-m_q(\mu))}, \non \\
 &&  r^{a_0}_\chi(\mu)={2m_{a_0}^2\over
 m_b(\mu)(m_d(\mu)-m_u(\mu))}, \quad \bar r_\chi^{a_0}(\mu)={2m_{a_0}\over
 m_b(\mu)},  \quad \bar r_\chi^{f_0}(\mu)={2m_{f_0}\over m_b(\mu)}.
 \en
Note that the $f_0$--$\sigma$ mixing angle (i.e. $\sin\theta$) and
Clebsch-Gordon coefficient $1/\sqrt2$ have been included in the
$f_0(980)$ form factors $F^{Bf^{u,d}_0}$ and decay constants
$f_{f_0}^{u,d}$ and likewise for the form factors
$F^{B\sigma^{u,d}_0}$ and decay constants $f_{\sigma}^{u,d}$.
Throughout, the order of the arguments of the $a_i^p(M_1M_2)$ and
$b_i(M_1M_2)$ coefficients is dictated by the subscript $M_1M_2$,
where $M_2$ is the emitted meson and $M_1$ shares the same
spectator quark with the $B$ meson. For the annihilation diagram,
$M_1$ is referred to the one containing an antiquark from the weak
vertex, while $M_2$ contains a quark from the weak vertex.

\section{Determination of the scalar couplings of scalar mesons}
To determine the scalar decay constant $\bar f_S$ of the scalar
meson $S$ defined by $\langle 0| \bar q_2 q_1 |S \rangle =m_{S}
\bar f_S$, we consider the following two-point correlation
function
\begin{eqnarray}\label{app:masssrf0}
\Pi (q^2)=i\int d^4x e^{iqx} \langle 0|{\rm T} (j^{q_2q_1}(x)
j^{{q_2q_1}\dag} (0)|0\rangle \,,
\end{eqnarray}
with $j^{q_2q_1}=\bar q_2 q_1$. The above correlation function can
be calculated from the hadron and quark-gluon dynamical points of
view, respectively. Therefore, the correlation function arising
from the lowest-lying meson $S$ can be approximately
 written as
\begin{eqnarray}
\frac{m_S^2 \bar f_S^2}{m_S^2-q^2}= \frac{1}{\pi}\int^{s_0}_0 ds
\frac{{\rm Im} \Pi^{\rm OPE}}{s-q^2} \,,
\label{eq:higherresonance}
\end{eqnarray}
where $\Pi^{\rm OPE}$ is the QCD operator-product-expansion (OPE)
result at the quark-gluon level, $s_0$ is the threshold of the
higher resonant states, and the contributions originating from
higher resonances are approximated by
\begin{eqnarray}
\frac{1}{\pi}\int_{s_0}^\infty ds \frac{{\rm Im} \Pi^{\rm
OPE}}{s-q^2}\,.
\end{eqnarray}
We apply the Borel transformation to both sides of
Eq.~(\ref{eq:higherresonance}) to improve the convergence of the
OPE series and suppress the contributions from higher resonances.
Consequently, the sum rule for lowest lying resonance with OPE
series up to dimension 6 and ${\cal O}(\alpha_s)$ corrections
reads~\cite{GRVW}
\begin{eqnarray}
&& m_{S}^{2} \bar f_S^2 e^{-m_{S}^{2}/M^2}\Bigg
(\frac{\alpha_s(\mu)}{\alpha_s(M)}\Bigg)^{8/b} = \frac{3}{8\pi^2}
M^4 \Biggl[ 1 + \frac{\alpha_s(M)}{\pi}\Bigg( \frac{17}{3} +2
\frac{I(1)}{f(1)}-2\ln \frac{M^2}{\mu^2} \Bigg)f(1) \Biggr]
\nonumber\\
&&\ \ + \frac{1}{8}\langle \frac{\alpha_s G^2}{\pi}\rangle +
 \bigg(\frac{1}{2}m_1+m_2\bigg) \langle \bar q_1 q_1\rangle +
 \bigg(\frac{1}{2}m_2+m_1\bigg) \langle \bar q_2 q_2\rangle
 \nonumber\\
 &&\ \
 -\frac{1}{M^2}\Bigg(
  \frac{1}{2}m_2 \langle \bar q_1 g_s \sigma\cdot G q_1\rangle +
  \frac{1}{2}m_1 \langle \bar q_2 g_s \sigma\cdot G q_2\rangle
  - \pi\alpha_s\langle\bar q_1\sigma_{\mu\nu}\lambda^a q_2 \,
 \bar q_2\sigma^{\mu\nu}\lambda^a q_1\rangle
 \nonumber\\
  &&\ \ \ \ - \pi\alpha_s
 \frac{1}{3}\langle\bar q_1\gamma_\mu \lambda^a q_1 \,
 \bar q\gamma^\mu \lambda^a q_1\rangle - \pi\alpha_s
\frac{1}{3}\langle\bar q_2\gamma_\mu \lambda^a q_2 \,
 \bar q_2\gamma^\mu \lambda^a q_2\rangle \Bigg) \,,
\label{app:masssr}
\end{eqnarray}
where $f(1)=1-e^{-s_0/M^2}(1+s_0/M^2)$,
$I(1)=\int^1_{e^{-s_0/M^2}}(\ln t) \ln (-\ln t)dt$, the scale
dependence of  $\bar f_S$ is
\begin{eqnarray}
\bar f_S (M)=\bar f_S(\mu)\Bigg( \frac{\alpha_s(\mu)}{\alpha_s(M)}
\Bigg)^{4/b},
\end{eqnarray}
and the anomalous dimensions of relevant operators can be found in
Ref.~\cite{Yang:1993bp} to be
 \begin{eqnarray}\label{app:anamolous}
 && m_{q, \mu}= m_{q , \mu_0}
  \left(\frac{\alpha_s(\mu_0)}{\alpha_s(\mu)}\right)^{-{4\over b}},
  \nonumber\\
 &&\langle \bar q q\rangle_\mu = \langle \bar q q\rangle_{\mu_0}
 \left(\frac{\alpha_s(\mu_0)}{\alpha_s(\mu)}\right)^{4\over b},\nonumber\\
 && \langle g_s \bar q\sigma\cdot G q\rangle_\mu =
 \langle g_s \bar q\sigma\cdot G q\rangle_{\mu_0}
 \left(\frac{\alpha_s(\mu_0)}{\alpha_s(\mu)}\right)^{-{2\over 3b}},\nonumber\\
 && \langle \alpha_s G^2\rangle_\mu = \langle \alpha_s G^2\rangle_{\mu_0}
 ,
 \end{eqnarray}
with $b=(11 N_c -2n_f)/3$, where we have neglected the anomalous
dimensions of the 4-quark operators. In the numerical analysis, we
shall use $\alpha_s(1~{\rm GeV})=0.517$ corresponding to the world
average $\alpha_s(m_Z)=0.1213$, and the following values for
vacuum condensates and quark masses at the scale
$\mu=1$~GeV~\cite{Yang:1993bp}:
\begin{eqnarray}
\begin{array}{lcl}
  \langle \alpha_s G_{\mu\nu}^a G^{a\mu\nu} \rangle=0.474\ {\rm GeV}^4/(4\pi)\,, &  &   \\
  \langle \bar uu \rangle \cong \langle \bar dd \rangle =-(0.24\ {\rm
  GeV})^3 \,,
  &   & \langle \bar ss \rangle = 0.8 \langle \bar uu \rangle \,, \\
  (m_u+m_d)/2=5\ {\rm MeV}\,, &  & m_s=119\ {\rm MeV}\,,\\
  \langle g_s \bar u\sigma Gu \rangle \cong \langle g_s\bar
d\sigma Gd \rangle =-0.8\langle \bar uu \rangle, &  &\langle g_s
\bar s\sigma Gs \rangle = 0.8 \langle g_s\bar u\sigma Gu \rangle.
\end{array}\label{eq:parameters}
 \end{eqnarray}
We adopt the vacuum saturation approximation for describing the
four-quark condensates, i.e.,
\begin{eqnarray}
\langle 0|\bar q \Gamma_i \lambda^a q \bar q \Gamma_i \lambda^a
q|0\rangle =-\frac{1}{16N_c^2}{\rm Tr}(\Gamma_i\Gamma_i) {\rm
Tr}(\lambda^a \lambda^a) \langle \bar qq\rangle^2 \,.
\end{eqnarray}
Taking the logarithm of both sides of Eq.~(\ref{app:masssr}) and
then applying the differential operator $M^4 \partial /\partial
M^2$ to them, one can obtain the mass sum rule for the
lowest-lying resonance $S$, where $s_0$ is determined by the
maximum stability of the sum rule. Substituting the obtained $s_0$
and mass into Eq.~(\ref{app:masssr}), one arrives at the sum rule
for the decay constant $\bar f_S$.

 Nevertheless, in order to extract the decay constant $\bar f_{S'}$ for the first excited
 state $S'$, we shall consider two lowest lying
 states on the left hand side of Eq.~(\ref{app:masssr}), i.e.,
\begin{eqnarray}
 m_S^2 \bar f_S^2\, e^{-m_S^2/M^2}
 +m_{S'}^2 \bar f_{S'}^2\, e^{-m_{S'}^2/M^2}=
\frac{1}{\pi}\int^{s_0^\prime}_0 ds\, e^{-s/M^2}\, {\rm Im}
\Pi^{\rm OPE}(s) \,. \label{app:firstexcitedstate}
\end{eqnarray}

\subsection{$a_0(980)$ and $a_0(1450)$}

Taking $\bar q_1 q_2 \equiv \bar u d$ and considering only the
ground state meson, we obtain
\begin{eqnarray}\label{app:massa0980}
&& m_{S}\simeq (0.99\pm 0.05)~{\rm GeV}, \nonumber\\
&& \bar f_S(1~{\rm GeV})= 370~{\rm MeV}, \qquad
 \bar f_S(2.1~{\rm GeV})= 440~{\rm MeV}\,,
\end{eqnarray}
corresponding to $s_0\simeq 3.1~{\rm GeV}^2$ and the Borel window
$1.1~{\rm GeV}^2 <M^2< 1.6$~GeV$^2$, so that the resulting mass is
consistent with $a_0(980)$. However, if one would like to have the
mass result of the ground state to be consistent with that of
$a_0(1450)$, then one should choose a larger $s_0\simeq
6.0$~GeV$^2$ together with the Borel window with a larger
magnitude: $2.6~{\rm GeV}^2 <M^2< 3.1$~GeV$^2$. Since $\kappa,
a_0(980)$ and $f_0(980)$ may be four-quark states, we therefore
explore two possible scenarios: (i) In scenario 1, we treat
$\kappa, a_0(980), f_0(980)$ as the lowest lying states, and
$K_0^*(1430), a_0(1450), f_0(1500)$ as the corresponding first
excited states, respectively, where we have assumed that
$f_0(980)$ and $f_0(1500)$ are dominated by the $\bar s s$
component and (ii) we assume in scenario 2 that $K_0^*(1430),
a_0(1450), f_0(1500)$ are the lowest lying resonances and the
corresponding first excited states lie between $(2.0\sim
2.3)$~GeV. Scenario 2 corresponds to the case that light scalar
mesons are four-quark bound states, while all scalar mesons are
made of two quarks in scenario 1.

In the numerical analysis, we adopt the first two lowest
resonances as inputs in these two scenarios and perform the
quadratic fits to both the left-hand side and right-hand side of
the renormalization-improved sum rules in
Eq.~(\ref{app:firstexcitedstate}). We find that in scenario 1 the
resulting threshold and Borel window are $s_0=(5.0\pm
0.3)$~GeV$^2$ and $1.1~{\rm GeV}^2 <M^2< 1.6$~GeV$^2$,
respectively, while in scenario 2, $s_0=(9.0\pm 1.0)$~GeV$^2$ and
$2.6~{\rm GeV}^2 <M^2< 3.1$~GeV$^2$. Thus for $a_0(980)$ and
$a_0(1450)$, we obtain
\begin{equation}
 \begin{array}{ll}\label{app:decayconstantsa01}
 \bar f_{a_0(980)}(1~{\rm GeV})= (365\pm 20)~{\rm MeV},\ \ &
 \bar f_{a_0(980)}(2.1~{\rm GeV})= (450\pm 25)~{\rm MeV}, \\
 \bar f_{a_0(1450)}(1~{\rm GeV})= -(280\pm 30)~{\rm MeV}, &
 \bar f_{a_0(1450)}(2.1~{\rm GeV})= -(345\pm 35)~{\rm
 MeV},
 \end{array}
 \end{equation}
in scenario 1 and
\begin{eqnarray}\label{app:decayconstantsa02}
 \begin{array}{ll}
 \bar f_{a_0(1450)}(1~{\rm GeV})= (460\pm 50)~{\rm MeV},  \ \ &
 \bar f_{a_0(1450)}(2.1~{\rm GeV})= (570\pm 60)~{\rm MeV}, \\
 \bar f_{S'}(1~{\rm GeV})= (390\pm 80)~{\rm MeV}, &
 \bar f_{S'}(2.1~{\rm GeV})= (480\pm 100)~{\rm MeV},
 \end{array}
\end{eqnarray}
in scenario 2, where $S'$ denotes the first excited state. Note
that the sign of the decay constants for the excited states in
scenario 1 cannot be determined in the QCD sum rule approach [see
Eqs. (B.9) and (C.6)]. They are fixed from the signs of the form
factors as shown in Table~\ref{tab:FF} using the potential model
calculation.~\footnote{In the quark model with a simple harmonic
like potential, the wave functions for a state with the quantum
numbers $(n,l,m)$ is given by $f_{nl}(\vec p^2/\beta^2)
Y_{lm}(\hat p) \exp(-\vec p^2/2\beta^2)$ up to an overall sign,
with $f_{10}(x)=f_{11}(x)=1$ and $f_{21}(x)=\sqrt{5/2}(1-2x/5)$.
For the $n=2,l=1$ state, the decay constant $\bar f_S$ is
dominated by the second term in $f_{21}$, while  the $B\to S$ form
factors is governed by the first term in $f_{21}$ as the spectator
light quark in the $B$ meson is soft. Consequently, the decay
constant and the form factor for the excited state have opposite
signs. The overall sign with the wave function can be fixed by the
sign of the form factor which is chosen to be positive in general
practice.}

\subsection{$f_0(980)$ and $f_0(1500)$}

Here we will assume that $f_0(980)$ and $f_0(1500)$ are both
dominated by the $\bar s s$ component, i.e. $j^{s s}=\bar s s$.
The results read
\begin{eqnarray}\label{app:decayconstantsf01}
 \begin{array}{ll}
 \bar f_{f_0(980)}(1~{\rm GeV})= (370\pm 20)~{\rm MeV},  \ \ &
 \bar f_{f_0(980)}(2.1~{\rm GeV})= (460\pm 25)~{\rm MeV}, \\
 \bar f_{f_0(1500)}(1~{\rm GeV})= -(255\pm 30)~{\rm MeV}, &
 \bar f_{f_0(1500)}(2.1~{\rm GeV})= -(315\pm 35)~{\rm MeV},
 \end{array}
\end{eqnarray}
in  scenario 1, and
\begin{eqnarray}\label{app:decayconstantsf02}
 \begin{array}{ll}
 \bar f_{f_0(1500)}(1~{\rm GeV})= (490\pm 50){\rm MeV}, \ \ &
 \bar f_{f_0(1500)}(2.1~{\rm GeV})= (605\pm 60)~{\rm MeV}, \\
 \bar f_{S'}(1~{\rm GeV})= (375\pm 80)~{\rm MeV}, &
 \bar f_{S'}(2.1~{\rm GeV})= (465\pm 100)~{\rm MeV},
 \end{array}
\end{eqnarray}
in scenario 2.

\subsection{$\kappa(800)$ and $K_0^*(1430)$}

The relevant current is $j^{q s}=\bar q s$ with $\bar q= \bar u$
or $\bar d$ for the cases of $\kappa(800)$ and $K_0^*(1430)$.
Using the single resonance approximation as given in
Eq.~(\ref{app:masssr}), we find that the lowest lying mass roughly
equals to $(0.86\pm 0.02)$~GeV$^2$, corresponding to $s_0\simeq
2.4$~GeV and the Borel window of $0.8~{\rm GeV}^2 <M^2<
1.3$~GeV$^2$. In analogy with the case of $a_0(1450)$, if
$K_0^*(1430)$ is justified by the result of the lowest lying mass
sum rule, then it is necessary to have a large threshold
$s_0\simeq 6.0$~GeV$^2$ corresponding to a larger Borel mass
region $2.6~{\rm GeV}^2 <M^2< 3.1$~GeV$^2$, where the stable
plateau can be reached.

For $\kappa(800)$ and $K_0^*(1430)$, we find
\begin{eqnarray}\label{app:decayconstantsK01}
 \begin{array}{ll}
 \bar f_{\kappa(800)}(1~{\rm GeV})= (340\pm 20)~{\rm MeV}, \ \ &
 \bar f_{\kappa(800)}(2.1~{\rm GeV})= (420\pm 25)~{\rm MeV}, \\
 \bar f_{K_0^*(1430)}(1~{\rm GeV})= -(300\pm 30)~{\rm MeV}, &
 \bar f_{K_0^*(1430)}(\mu=2.1~{\rm GeV})= -(370\pm 35)~{\rm
 MeV},
 \end{array}
\end{eqnarray}
in scenario 1 and
\begin{eqnarray} \label{app:decayconstantsK02}
 \begin{array}{ll}
 \bar f_{K_0^*(1430)}(1~{\rm GeV})= (445\pm 50)~{\rm MeV}, \ \ &
 \bar f_{K_0^*(1430)}(2.1~{\rm GeV})= (550\pm 60)~{\rm MeV}, \\
 \bar f_{S'}(1~{\rm GeV})= -(420\pm 80)~{\rm MeV}, &
 \bar f_{S'}(2.1~{\rm GeV})= -(520\pm 100)~{\rm MeV},
 \end{array}
\end{eqnarray}
in scenario 2.

Two remarks are in order. First, if neglecting the RG improvement
for the mass sum rules and considering only the lowest lying
resonance state, the results, as stressed in Ref.~\cite{GRVW},
become sensitive to the values of the four quark condensates for
which the vacuum saturation approximation has been applied. Then
it is possible to have results to be consistent $a_0(1450),
K_0^*(1430)$ and $f_0(1500)$ in the range of 0.8~GeV$^2 < M^2 <1.2
$~GeV$^2$ if $s_0$ is larger and four-quark condensates are
several times larger than that in the vacuum saturation
approximation. Second, thus far we have considered
renormalization-group (RG) improved QCD sum rules. It is found
that sum rule results become insensitive to four-quark condensates
if the RG improved effects are considered. For the RG improved
mass sum rules, if taking $a_0(1450), K_0^*(1430)$ and $f_0(1500)$
as lowest resonances, then it is necessary to have a large
threshold $s_0\gtrsim 4.9$~GeV$^2$ corresponding to a much larger
Borel mass region $2.6~{\rm GeV}^2 <M^2< 3.2$~GeV$^2$, in contrast
with the conclusion in Ref.~\cite{Du} where the stable Borel
window for the $K_0^*(1430)$ mass sum rule is $1.0~{\rm GeV}^2
<M^2< 1.2$~GeV$^2$.

\section{Leading twist LCDAs for scalar mesons}
The LCDA $\Phi_{S}(x,\mu)$ corresponding to the quark content
$q_1\bar q_2$ is defined by
 \be
 \la S(p)|\bar q_1(z)\gamma_\mu q_2(0)|0\ra
 &=& p_\mu\int^1_0 dx e^{ixp\cdot z}\Phi_{S}(x,\mu),
 \en
where $x$ ($\bar x=1-x$) is the momentum fraction carried by the
quark $q$ (antiquark $\bar q$) and $\mu$ is the normalization
scale of the LCDA. $\Phi_{S}(x,\mu)$ can be expanded in a series
of Gegenbauer polynomials~\cite{Chernyak:1983ej,Braun}
\begin{eqnarray}\label{app:generalda}
\Phi_{S}(x,\mu)=\bar f_S 6x(1-x)\Bigg[\sum_{l=0}^\infty B_l (\mu)
C^{3/2}_l(2x-1)\Bigg],
\end{eqnarray}
where multiplicatively renormalizable coefficients (or the
so-called Gegenbauer moments) are given by
\begin{eqnarray} \label{eq:Bm}
B_l(\mu) = \frac{1}{\bar f_S} \frac{2(2l+3)}{3(l+1)(l+2)} \int_0^1
C^{3/2}_l (2x-1) \Phi_{S}(x,\mu)\, dx ,
\end{eqnarray}
which vanish for even $l$ in the SU(3) limit. Consider the
following two-point correlation function
\begin{eqnarray}
\Pi_l (q) = i\int d^4x e^{iqx} \langle 0| T(O_l(x)\, O^\dagger(0)
|0 \rangle = (zq)^{l+1} I_l (q^2),
\end{eqnarray}
where
\begin{eqnarray}
&&\langle 0| O_l|S(p)\rangle \equiv \langle 0| \bar q_2 \not\! z
(i z \stackrel{\leftrightarrow}{D})^l q_1 |S(p)\rangle=(zp)^{l+1}
  \int^1_0 (2x-1)^l
\Phi_{S}(x) dx \equiv (zp)^{l+1}  \bar f_S \langle \xi^l_{S}
\rangle,  \non
 \hspace{1cm}\\
&& \langle 0| O|S(p)\rangle \equiv \langle 0| \bar q_2 q_1
|S(p)\rangle = m_{S} \bar{f}_S,
\end{eqnarray}
with $z^2=0$ and $\xi=2x-1$.

We shall saturate the physical spectrum with two lowest lying
resonances for reasons to be explained later. Therefore, the
correlation function $I_l$ can be approximately written as
\begin{eqnarray}
\frac{m_S^2 \bar f_S^2 \langle \xi^l_S\rangle}{m_S^2-q^2}+
 \frac{m_{S'}^2 \bar f_{S'}^2 \langle \xi^l_{S'}\rangle}{m_{S'}^2-q^2}=
\frac{1}{\pi}\int^{s_0}_0 ds \frac{{\rm Im} I_l^{\rm OPE }
(s)}{s-q^2} \,, \label{eq:}
\end{eqnarray}
where $S$ and $S'$ refer to the lowest and first excited resonance
states, respectively, and
 \begin{eqnarray}\label{app:srmoments1}
I_l(q^2) &=&
 \frac{3 }{16\pi^2} \Bigg(\frac{m_{q_2}
 + m_{q_1}}{l+2} + \frac{m_{q_2} - m_{q_1}}{l+1} \Bigg)
 \ln\Bigg(\frac{-q^2}{\mu^2}\Bigg)
 - \frac{\langle \bar q_2 q_2\rangle}{q^2} +
 \frac{10l-3}{24} \frac{\langle \bar q_2 g_s \sigma\cdot G q_2
 \rangle}{q^4} \nonumber\\
 && \ \ \
 -  \frac{l(4l-5)}{18}{\langle g_s^2 G^2\rangle \langle \bar q_2 q_2\rangle \over q^6}
  +  (-1)^{l+1}  \Bigg[ \frac{3 }{16\pi^2 } \Bigg(\frac{m_{q_2}
+ m_{q_1}}{l+2} - \frac{m_{q_2} - m_{q_1}}{l+1} \Bigg)
 \ln\Bigg(\frac{-q^2}{\mu^2}\Bigg)
 \nonumber\\  && \ \ \  -\frac{\langle \bar q_1 q_1\rangle}{q^2}  +
 \frac{10l-3}{24} \frac{\langle \bar q_1 g_s \sigma\cdot G q_1 \rangle}{q^4}
 - \frac{l(4l-5)}{18}{\langle g_s^2 G^2\rangle \langle \bar q_1 q_1\rangle \over q^6}
 \Bigg] .
 \end{eqnarray}
In terms of the above defined moments $\langle \xi^l_S \rangle$,
the sum rule reads
 \begin{eqnarray}\label{app:srmoments2}
 &&\langle \xi^l_S\rangle {m_S \bar f_S^2}
e^{-m_S^2/M^2} + \langle \xi^l_{S'}\rangle {m_{S'} \bar f_{S'}^2}
e^{-m_{S'}^2/M^2} \nonumber\\
&&= \Bigg\{ -\frac{3 }{16\pi^2 }M^2 \Bigg(\frac{m_{q_2}
 + m_{q_1}}{l+2} + \frac{m_{q_2} - m_{q_1}}{l+1} \Bigg)f(0)
 + \langle \bar q_2 q_2\rangle +
 \frac{10l-3}{24} \frac{\langle \bar q_2 g_s \sigma\cdot G q_2
 \rangle}{M^2} \nonumber\\
 && \ \ \
 +  \frac{l(4l-5)}{36}{\langle g_s^2 G^2\rangle \langle \bar q_2 q_2\rangle \over M^4}
  +  (-1)^{l+1}  \Bigg[ -\frac{3 }{16\pi^2 }M^2 \Bigg(\frac{m_{q_2}
+ m_{q_1}}{l+2} - \frac{m_{q_2} - m_{q_1}}{l+1} \Bigg)f(0)
 \nonumber\\  && \ \ \  + \langle \bar q_1 q_1\rangle  +
 \frac{10l-3}{24} \frac{\langle \bar q_1 g_s \sigma\cdot G q_1 \rangle}{M^2}
 + \frac{l(4l-5)}{36}{\langle g_s^2 G^2\rangle \langle \bar q_1 q_1\rangle \over M^4}
 \Bigg]\Bigg\},
 \end{eqnarray}
with $f(0)=1-e^{-s_0/M^2}$, while the Gegenbauer moments are given
by
\begin{eqnarray}
B_l^{S^{(\prime)}}(\mu) = \frac{1}{\bar f_{S^{(\prime)}}}
\frac{2(2l+3)}{3(l+1)(l+2)} \langle C^{3/2}_l (\xi_S^{(\prime)})
\rangle.
\end{eqnarray}
Conformal invariance in QCD indicates that partial waves in the
expansion of $\Phi_S(x,\mu)$ in Eq.~(\ref{app:generalda}) with
different conformal spin cannot mix under renormalization to the
leading-order accuracy. Consequently, the Gegenbauer moments $B_l$
renormalize multiplicatively:
  \begin{equation}
    B_l(\mu) = B_l(\mu_0)
  \left(\frac{\alpha_s(\mu_0)}{\alpha_s(\mu)}\right)^{-(\gamma_{(l)}+4)/{b}},
  \label{momentdar}
   \end{equation}
where the one-loop anomalous dimensions are \cite{GW}
  \begin{eqnarray}
  \gamma_{(l)}  = C_F
  \left(1-\frac{2}{(l+1)(l+2)}+4 \sum_{j=2}^{l+1} \frac{1}{j}\right),
  \label{eq:1loopandim}
  \end{eqnarray}
with $C_F=(N_c^2-1)/(2N_c )$. Note that $\bar f_S B_0$ is
independent of the renormalization scale. It should be also
stressed that if only the lowest resonances are taken into account
in Eq.~(\ref{app:srmoments2}), the resultant mass reading from the
sum rule that follows the same line as before by taking $[(M^4
\partial /\partial M^2)\, \ln ]$ to both sides of
Eq.~(\ref{app:srmoments2}) is less than $0.4$~GeV, which is too
small compared with the observables. Therefore, in the numerical
analysis, we shall consider the first two lowest resonances and
perform the quadratic fits to both the left-hand side and
right-hand side of the renormalization-improved moment sum rules,
given in Eq.~(\ref{app:srmoments2}), within the Borel window
$M_{\min}< M^2 < M_{\max}^2$ with $M_{\min}^2, M_{\max}^2 \in
(1.1~{\rm GeV}^2,1.6~{\rm GeV}^2)$ [and $M_{\min}^2, M_{\max}^2
\in (0.8~{\rm GeV}^2,1.3~{\rm GeV}^2)$] corresponding to
$\langle\xi_{a_0, f_0}\rangle$ [and
$\langle\xi_{\kappa,K_0^*(1430}\rangle$] in scenario 1 and
$M_{\min}^2, M_{\max}^2 \in (2.6~{\rm GeV}^2,3.1~{\rm GeV}^2)$ in
scenario 2, where the Borel windows are same as those in the
previous section. It should be noted that for the moment sum rule
for $\langle\xi^l\rangle$ in the large $l$ limit, the actual
expansion parameter is $M^2/l$. Therefore, for
$\langle\xi^3\rangle$ we rescale the Borel windows to be
$M_{\min}^2, M_{\max}^2 \in (1.4~{\rm GeV}^2,1.9~{\rm GeV}^2)$ for
$a_0, f_0$ [and $M_{\min}^2, M_{\max}^2 \in (1.1~{\rm
GeV}^2,1.6~{\rm GeV}^2)$ for $\kappa, K_0^*(1430)$] in scenario 1
and $M_{\min}^2, M_{\max}^2 \in (2.9~{\rm GeV}^2,3.4~{\rm GeV}^2)$
in scenario 2. Furthermore, for $l\geq 5$ and fixed $M^2$, the OPE
series are convergent slowly or even divergent, i.e. the resulting
sum-rule result becomes less reliable. Following the same line as
given in the previous section, we explore two possible scenarios.
The results for the fist and second moments of $\la \xi^l\ra$
together with the fist and second Gegenbauer moments are collected
in Tables~\ref{tab:momentscenario1} and \ref{tab:momentscenario2}.
\begin{table}[htb]
\caption{Gegenbauer moments at the scales $\mu=1$ GeV and 2.1 GeV
(shown in parentheses) in scenario 1. }
\label{tab:momentscenario1}
\begin{ruledtabular}
\begin{tabular}{c|rr|rr}
 State & $\langle \xi \rangle$ &  $\langle \xi^3 \rangle$
 & $B_1$ & $B_3$ \\ \hline
 $a_0(980)$ & $-0.56\pm 0.05$ &  $-0.21\pm 0.03$
  & $-0.93\pm 0.10 (-0.64\pm 0.07)$ & $0.14\pm 0.08\ \ (0.08\pm 0.04)$      \\ \hline
 $a_0(1450)$ & $0.53\pm 0.20$ &  $0.00\pm 0.04$
  & $0.89\pm 0.20\ \  (0.62\pm 0.14)$ & $-1.38\pm0.18 (-0.81\pm0.11)$  \\ \hline\hline
 $f_0(980)$ & $-0.47\pm0.05$ &  $-0.20\pm 0.03$
  & $-0.78\pm0.08 (-0.54\pm 0.06)$ & $0.02\pm 0.07\ \ (0.01\pm 0.04)$      \\ \hline
 $f_0(1500)$ & $0.48\pm0.24$ &  $-0.05\pm 0.04$
  & $0.80\pm 0.40\ \ (0.47\pm 0.28)$ & $-1.32\pm 0.14 (-0.77\pm 0.08)$  \\ \hline\hline
 $\kappa(800)$ & $-0.55\pm 0.07$ & $-0.21\pm 0.05$
  & $-0.92\pm 0.11 (-0.64\pm 0.08)$ & $0.15\pm 0.09\ \  (0.09\pm 0.05)$  \\ \hline
 $K_0^*(1430)$ & $0.35\pm 0.07$ & $-0.08\pm 0.06$
  & $0.58\pm 0.07\ \  (0.39\pm 0.05)$ & $-1.20\pm 0.08 (-0.70\pm 0.05)$      \\
\end{tabular}
\end{ruledtabular}
\end{table}

\begin{table}[htb]
\caption{Same as Table \ref{tab:momentscenario1} except for
scenario 2.} \label{tab:momentscenario2}
\begin{ruledtabular}
\begin{tabular}{c|rr|rr}
 State & $\langle \xi \rangle$ &  $\langle \xi^3 \rangle$
 & $B_1$ & $B_3$ \\ \hline
 $a_0(1450)$ & $-0.35\pm0.07$ &  $-0.24\pm 0.06$
  & $-0.58\pm 0.12 (-0.40\pm 0.08)$ & $-0.49\pm 0.15 (-0.29\pm 0.09)$  \\ \hline
 higher resonance & $0.44\pm 0.27$ & $0.22\pm 0.11$
  & $0.73\pm 0.45\ \ (0.51\pm 0.26)$ & $0.17\pm0.20\ \ (0.10\pm 0.12)$  \\ \hline\hline
 $f_0(1500)$ & $-0.29\pm0.06$ &  $-0.19\pm 0.05$
  & $-0.48\pm 0.11 (-0.33\pm 0.08)$ & $-0.37\pm 0.20 (-0.22\pm 0.12)$  \\ \hline
 higher resonance & $0.34\pm0.30$ & $0.16\pm 0.15$
  & $0.56\pm 0.50\ \ (0.39\pm 0.35)$ & $0.07\pm 0.23\ \ (0.04\pm 0.13)$  \\ \hline\hline
 $K_0^*(1430)$ & $-0.35\pm 0.08$ & $-0.23\pm 0.06$
  & $-0.57\pm 0.13 (-0.39\pm 0.09)$ & $-0.42\pm 0.22 (-0.25\pm 0.13)$   \\ \hline
 higher resonance &  $0.25\pm 0.11$ & $0.12\pm 0.05$
  & $0.41\pm 0.34\ \ (0.28\pm 0.24)$ & $0.09\pm 0.14\ \ (0.05\pm 0.08)$     \\
\end{tabular}
\end{ruledtabular}
\end{table}


\end{document}